\documentclass[12pt]{JHEP3}
\usepackage{epsfig}
\usepackage{amsmath}
\newcommand{\unit}{\hbox to 3.8pt{\hskip1.3pt \vrule height 7.4pt
    width .4pt \hskip.7pt \vrule height 7.85pt width .4pt \kern-2.4pt 
    \hrulefill \kern-3pt \raise 3.7pt\hbox{\char'40}}}
\newcommand{\p}{\partial}
\newcommand{\ket}[1]{{\left|#1\right\rangle}}
\newcommand{\bra}[1]{{\left\langle#1\right|}}
\title{
ADHM is Tachyon Condensation
}
\author{
Koji Hashimoto$^\dagger$ and Seiji Terashima$^*$\\ 
\hspace{-2mm}${}^\dagger$ {\it Institute of Physics, 
University of Tokyo, Komaba 3-8-1, Tokyo 153-8902, Japan}\\
{\it Kavli Institute for Theoretical Physics,
University of California,}\\
\hspace{3mm} {\it Santa Barbara, CA 93106-4030, U.S.A.}\\
{\it DAMTP, University of Cambridge,
Wilberforce Rd, Cambridge CB30WA, U.K.}\\
E-mail: \email{koji@hep1.c.u-tokyo.ac.jp}\\
\hspace{-2mm}${}^*$ 
{\it New High Energy Theory Center, Rutgers University,
126 Frelinghuysen Road,}\\
\hspace{3mm} {\it Piscataway, NJ 08854-8019, U.S.A.}\\
E-mail: \email{seijit@physics.rutgers.edu}\\
}

\abstract{
We completely realize the ADHM construction of instantons in D-brane
language of tachyon condensations. 
Every step of the construction is given
a physical interpretation in string theory, in  
a boundary state formalism valid all order in $\alpha'$.  
Accordingly, equivalence between Yang-Mills configurations on D4-branes
and D0-branes inside the D4-branes is proven, which shows
that small instanton configurations of the Yang-Mills fields are
protected against stringy $\alpha'$ corrections. We provide also
D-brane realizations of the inverse ADHM construction, the completeness, 
and the noncommutative ADHM construction. 
}

\preprint{
{\normalsize{\tt hep-th/0511297}}\\ 
{\normalsize UT-Komaba/05-13}\\
{\normalsize DAMTP-2005-113}\\
{\normalsize NSF-KITP-05-96}
}

\begin{document}

\section{Introduction: Beyond $\alpha'$}
\label{sec:1}

Instantons are one of the most important nonperturbative effects in
field theories, and the tremendous success of string theory in
reproducing/predicting physical quantities in ordinary field theories
partially owes to the fact that these nonperturbative effects have 
counterpart in string theory, as D-branes within (or intersecting with)
D-branes. The most well-known example is the instantons in 4 dimensional
Yang-Mills theory, which are considered to be equivalent to a bound 
state of D0-branes and D4-branes
\cite{Witten:1995gx,Douglas:1995bn}. But how rigorously this equivalence
can be proven? 

The equivalence, so far, has been supported by various consistency
checks such as supersymmetries preserved, charges, masses and so 
on.\footnote{For a review of the D0-D4 system and the relevance to the
instantons, see \cite{Dorey:1999pd}. Instanton configurations are 
obtained from the D0-D4 system in \cite{Billo:2002hm}, but it is 
different from our standpoint: in \cite{Billo:2002hm} 
D0-branes were cosidered as a source for the Yang-Mills
fields.} 
Precisely speaking, the instanton (or self-dual) gauge fields
solve the equations of motion of Yang-Mills theory which is
the D4-brane world volume theory at low energy. 
Thus the instanton gauge configuration is trustable
only for large size of the instantons with which the $\alpha'$
corrections are negligible. 
On the other hand, the D0-branes
are point-like, and the description of them 
by the low energy effective action should be valid 
in a very different parameter region where the size of the
instanton is sub-stringy.\footnote{
The moduli space of a single instanton in SU$(2)$ Yang-Mills theory
consists of a real half-line specifying the instanton size $\rho$
(and ${\bf R}^4$ giving the location of the instanton).
The hypermultiplet appearing in the D0-brane effective action is 
related to the size as $S=\rho/\alpha'$,
which has mass dimension one. Thus the effective theory is valid for
$\alpha' S^2 \ll 1$, meaning $\rho \ll \sqrt{\alpha}$.}
Thus it is fair to say that
we do not know what is the configuration in string theory
corresponding to the low energy gauge theory instanton.  
It is possible that a configuration in a low energy effective theory 
would be extended (UV-completed) to a sting theory configuration 
in many different ways. 
This problem is not only for the instanton $\leftrightarrow$
D-brane equivalence but is rather generic in
arguments based on BPS properties (although they are 
useful in various situations), and is very difficult to solve
practically. Therefore it is quite interesting to connect these two
descriptions, the instantons and the D0-branes, explicitly. 
In this paper, we are going to show the equivalence of these two
pictures, beyond the $\alpha'$ corrections. 

A clue is hidden in the famous ADHM (Atiyah, Drinfeld, Hitchin and
Manin) construction of instantons \cite{Atiyah:1978ri}, in which
with ADHM data solving the ADHM equations an explicit self-dual
gauge field can be constructed.
Since the ADHM data have been identified with string excitations
connecting the D0-branes and the D4-branes \cite{Douglas:1995bn}, 
and the ADHM equations can be seen as BPS equations of the low energy
effective field theory on the D0-branes,
the ADHM construction explicitly relates these two
pictures.\footnote{
Note that the instanton equation and the ADHM equation are 
scale-invariant (in the commutative space-time).
Thus the moduli spaces near $\rho \ll \sqrt{\alpha'}$ 
and near $\rho \gg \sqrt{\alpha'}$
are of the same form, though these two regions are separated far way
from each other.
} 
However this provides a further question, because 
the valid regions of the descriptions are very different. 

In this paper, we will ``derive'' the ADHM construction from D-branes,
and realize all the procedures of the ADHM construction in a D-brane 
setup rigorously, and thus provide physical meaning for each procedure.
This implementation in terms of D-branes will be given in 
a boundary state formalism \cite{bs}
(and a boundary string field theory (BSFT) \cite{BSFT,ABS}),
instead of the low energy effective actions. Thus the ``derived'' ADHM
procedures are valid beyond stringy $\alpha'$ corrections,
which resolves the question above. In other words, on D-branes
the ADHM construction works regardless of the parameter regions in
concern. And, this shows the equivalence of the two descriptions
at all order in $\alpha'$.

Let us explain briefly how we derive the ADHM construction in string
theory which is valid at all order in $\alpha'$.
In string theory, different dimensional D-branes can be related via 
a K-theoretic argument \cite{WittenK} in which any kind of D-branes can
be obtained by a single kind of D-branes by {\it tachyon
condensation}. This is the D-brane descent \cite{senconj} /
ascent \cite{Terashima:2001jc} relations. 
For example, in two pairs of a D4-brane and an anti-D4-brane, 
condensation of the tachyon whose profile is linear in the worldvolume
coordinate leads to a single D0-brane 
\cite{Kraus:2000nj, Takayanagi:2000rz}
(Atiyah-Bott-Shapiro construction
\cite{WittenK, ABS}). In this way, the D0-branes
can be viewed as D4-branes, and along the way to include precisely 
the instantons into this scheme of the tachyon condensation,
surprisingly we find that the ADHM construction naturally emerges.
Therefore, {\it ADHM construction is nothing but a tachyon
condensation}. 

There have been attempts to realize the ADHM construction in D-branes
\cite{Douglas:1996uz,Diaconescu:1996rk,Hori:1999me}, 
but our rigorous equivalence provides not only the direct
relationship but also the following byproducts. The inverse ADHM
construction, with which for a given instanton configuration the ADHM
data is reproduced, can be derived in a similar manner as an ascent
relation of the tachyon condensation. Furthermore, in the (inverse)
ADHM construction, the completeness and the
uniqueness of the ADHM construction have been shown
\cite{Corrigan:1983sv}.  
This completeness
can be lifted to the D-brane language, which even provides a simple
proof of the completeness.
We can ``deconstruct'' any D-brane system by
infinitely many lowest dimensional D-branes and anti-D-branes.\footnote{ 
Instead of this, we can use a higher dimensional one, but
the lowest dimensional D-branes may be the simplest to study
\cite{kmatrix,Asakawa:2002ui}.} 
Thus there is a unified picture for any D-brane system. 
This underlies the realization of the completeness and the Nahm
construction of monopoles \cite{Nahm:1979yw} for which we gave a 
stringy realization in our previous paper 
\cite{Hashimoto:2005yy}. 

The organization of this paper is as follows. In Sec.~\ref{sec:2}, after
a review of the ADHM construction and the D0-D4 system, we explain our
idea of realizing the ADHM construction as a tachyon condensation.
Then the detailed proof is provided in Sec.~\ref{sec:3}, with a stringy 
derivation of the noncommutative ADHM construction
\cite{Nekrasov:1998ss}. Sec.~\ref{sec:4} is for the derivation of the
inverse ADHM construction and the completeness. In Sec.~\ref{sec:5},
according to the realization of the ADHM construction provided in this
paper, we give a conjecture stating that {\it the self-dual Yang-Mills 
configuration with arbitrary size solves non-Abelian Born-Infeld
equations of motion obtained in string theory to all order in $\alpha'$
including derivative corrections}.\footnote{
We show in this paper that the D0-D4 system possessing the ADHM data
without the constraint (the ADHM equation) provides the gauge fields 
on the D4-brane where the gauge field configurations are computed by
the ADHM construction. 
Therefore the equivalence is shown at off-shell.
(Precisely speaking, we need to require a weak condition on
the asymptotic behavior of the Dirac operators, but don't need the ADHM 
equation itself in showing the equivalence.)
In this sense the off-shell ADHM construction works in string
theory. Our result for on-shell configurations is at small instanton
singularity, which strongly supports this conjecture concerning
arbitrary points in the instanton moduli space. We note that similar
statements have been put within the context of ``Born-Infeld''
corrections \cite{Hashimoto:1997px}, but at the
best our our knowledge no statement including all the derivative
corrections has been made. 
(This might be subtle, in the sense that 
in the non-Abelian case the ``Born-Infeld corrections'' may not make 
sense because they can be traded with commutators of covariant
derivatives and the notion of ``constant'' field strength is not
well-defined.)}  
Seiberg and Witten \cite{Seiberg:1999vs} argued this from the viewpoint
of worldsheet supersymmetries.
Discussions in Sec.~\ref{sec:5} are
on various low-energy limits, the Atiyah-Singer index theorem
\cite{Atiyah:1968mp}, and generalizations of the ADHM construction.

\section{Tachyon Condensation and ADHM}
\label{sec:2}

\subsection{Review: ADHM construction of instantons and D0-D4 system}

Before explaining our strategy to derive the ADHM (and the inverse ADHM)
construction of instantons from the tachyon condensation of unstable
D-branes, we briefly summarize the ADHM construction 
itself and corresponding D-brane configurations in superstring theory.

The ADHM construction is a powerful tool for constructing gauge
configurations of instantons explicitly. For the construction of $k$
instanton configurations in SU$(N)$ Yang-Mills theory, we need
the following ADHM data: $S$ which is an $N\times 2k$ complex 
constant matrix and 
$X_\mu (\mu=1,2,3,4)$ which are hermitian 
$k\times k$ matrices. Then the
procedures of the ADHM construction starts with finding $N$ zeromodes
of a zero dimensional ``Dirac operator'' $\nabla^\dagger$, 
\begin{eqnarray}
 \nabla^\dagger V =0, \quad
\nabla^\dagger \equiv 
\bordermatrix{
& \overbrace{}^{N} 
& \overbrace{\hspace{25mm}}^{2k} \cr\nonumber\\[-8mm]
& S^\dagger & e^\dagger_\mu\! \otimes\! (x^\mu \unit_k- X^\mu)\!\!
\cr}
\Bigm\}
{\scriptstyle 2k}
\ .
\label{diraczero}
\end{eqnarray}
Here $e_\mu$ ($\mu=1,2,3,4$) are a representation of quaternion, 
$e_\mu \equiv (i\sigma_i, \unit_2)$, where $\sigma_i$ $(i=1,2,3)$
are Pauli matrices.
Arraying the $N$ independent zeromodes constitutes $V$ which is a 
$(N+2k)\times N$ matrix normalized as $V^\dagger V = \unit_N$. 
This $V$ is a function of $x^\mu$ through the Dirac operator 
$\nabla^\dagger$, and the desired instanton gauge field configuration is 
given by the formula
\begin{eqnarray}
 A_\mu = V^\dagger \p_\mu V \ .
\label{adhmformula}
\end{eqnarray}
For this gauge field to be self-dual, the ADHM data should satisfy
the ADHM equations
\begin{eqnarray}
 {\rm Tr}\left[
\sigma_i
(S^\dagger S + ((e^\dagger)^\mu e^\nu  X^\mu X^\nu) )
\right]=0 \quad (i=1,2,3)
\label{adhmeq}
\end{eqnarray}

In brane language, this system of instantons in Euclidean 4 dimensional
Yang-Mills theory has been known to be described by a combined brane
configuration of $k$ D0-branes and $N$ D4-branes in type IIA superstring
theory \cite{Douglas:1995bn}. The low energy effective field theory on
the $N$ D4-branes is the $1+4$ dimensional U$(N)$ Yang-Mills theory with
maximal supersymmetries. If we restrict our attention to the gauge
fields with spatial indices $A_\mu$ ($\mu=1,2,3,4$), then 
the instanton configurations equivalent to self-dual
configurations of the gauge fields are compatible with the BPS condition 
of preserving half of the supersymmetries on the worldvolume. 
The instanton charge is shown to be equal to the total D0-brane charge
bound on the D4-brane, through the Ramond-Ramond coupling in the
D4-brane action. Therefore, Yang-Mills instantons have the same amount
of charges, masses and supersymmetries, as those of the D0-branes on the
D4-branes.  

On the other hand, one can look at this brane system from the viewpoint
of the worldvolume effective field theory on the $k$ D0-branes. The low
energy matter content includes scalar field excitations $X^\mu$ from 
strings connecting the D0-branes, and another scalar field $S$
connecting the D0-branes and the D4-branes. The Chan-Paton factor
suggests that $X^\mu$ are $N\times N$ hermitian matrices, and $S$ is 
a complex $N \times k$ matrix tensored with an SU$(2)$ vector index
(this SU$(2)$ is one of SU(2)$\times$SU(2)$ \sim $SO(4) which is the
worldvolume Lorentz symmetry of the D4-branes and should be seen by the
D0-branes as a global symmetry, {\it i.e.}~the R-symmetry). 
This set ($S, X^\mu$) should be identified with the ADHM data
\cite{Douglas:1995bn}, and in fact, the BPS condition for ($S, X^\mu$)
is equivalent to the ADHM equation (\ref{adhmeq}).

These facts show that D-brane techniques are quite powerful in that
a part of the ingredients of the ADHM construction already
appear as matter contents and supersymmetry conditions on the
worldvolumes. 
Furthermore, introduction of small D-brane probes \cite{Douglas:1996uz}
enables one to actually realize the ADHM construction explained above.
In a D5-D9 system which is T-duality equivalent to the above D0-D4
system, one introduces a probe D1-brane whose effective worldvolume
sigma model realizes the ADHM formula (\ref{adhmformula}) for the
background gauge fields $A_\mu$ on the D9-branes. 
This probe analysis was generalized to the Nahm construction of monopole
\cite{Diaconescu:1996rk}
and the Nahm transformation for the gauge fields on $T^4$ 
by using the T-duality \cite{Hori:1999me}.

However, these interesting connections to the D-branes 
introduce probes, which means that one can get only the information
seen by the probes. Furthermore, the descriptions use effective actions
on the probe and so valid only in the low energy limits. 
Therefore, the probe method is not enough to show that 
in fact the two descriptions, one by the 
D4-branes (self-dual equations) and one by the D0-branes (ADHM
equations) are completely equivalent beyond the stringy corrections.
Another important point missing in the probe method is the inverse ADHM 
construction and the completeness \cite{Corrigan:1983sv}. 
The ADHM construction gives all the instanton configurations up to gauge
transformations, that is the completeness and the uniqueness of the
construction. This was shown explicitly \cite{Corrigan:1983sv}  
by applying the inverse ADHM construction to the gauge fields
constructed by the ADHM construction.\footnote{
The realization of the Nahm transform \cite{Hori:1999me} is 
showing the completeness, but there the worldvolume is compactified
and resultantly the information on the $S$ field is unclear.}

In the following, we present a complete derivation of the ADHM and the
inverse ADHM constructions, without using any probes, and in all order
in stringy corrections. This is possible owing to an exact 
treatment of the 
tachyon condensation in the BSFT and the
boundary state formalism.

\subsection{Our idea: brane configurations and tachyon condensation}
\label{secidea}
\setcounter{footnote}{0}

The powerfulness of the ADHM construction is due to the difference in
dimensions to solve. The ADHM equation is a purely algebraic equation
while the instanton equation is a partial differential equation which
is highly nontrivial. It is miraculous that those two are equivalent.
However, this miracle is shared by generic D-brane physics --- there is
a notion called ``brane democracy'' first mentioned by Townsend 
\cite{Townsend:1995gp} which is generalized to mean that through various
dualities any dimensional branes may play central role in constructing
the full string/M theory and in revealing 
dynamics of any other dimensional branes. One noble example 
is Matrix theory \cite{Banks:1996vh} 
in which lowest dimensional D-branes are
constituents to build higher dimensional M-theory physics.
A shortcoming of the Matrix theory is that charges of the constituents
remains in any setup made out of them, but it has been overcome by
K-matrix theory \cite{kmatrix,Asakawa:2002ui}
in which the constituents are unstable D-branes and without the
restriction of the charges 
one can truly construct any brane configurations out of them through
tachyon condensation \cite{senconj}, 
the annihilation of unstable D-branes, developed by Sen. 

The brane configuration of our concern consists of two different 
kinds of D-branes, the $k$ D0-branes and the $N$ D4-branes. 
In the sense described above, it is natural to consider a treatment of
this system in terms of a single kind of D-branes.  
There are two ways to realize this: 
\begin{itemize}
 \item[(a)] By D4-branes solely. One can represent the $k$ 
D0-branes by a tachyon condensation of $2k$ pairs of D4-branes and
anti-D4-branes. This is a D-brane descent relation. In total, one has 
$N+2k$ D4-branes and $2k$ anti-D4-branes.
\item[(b)] 
By D0-branes only. The $N$ D4-branes can be constructed by a tachyon
       condensation of infinite number of pairs
       of D0-branes and anti-D0-branes. This is called a D-brane ascent
       relation, found in \cite{Terashima:2001jc} and developed in 
       \cite{kmatrix,Asakawa:2002ui}.
\end{itemize}

\begin{figure}[t]
\begin{center}
 \begin{minipage}{7cm}
  \begin{center}
\includegraphics[width=7cm]{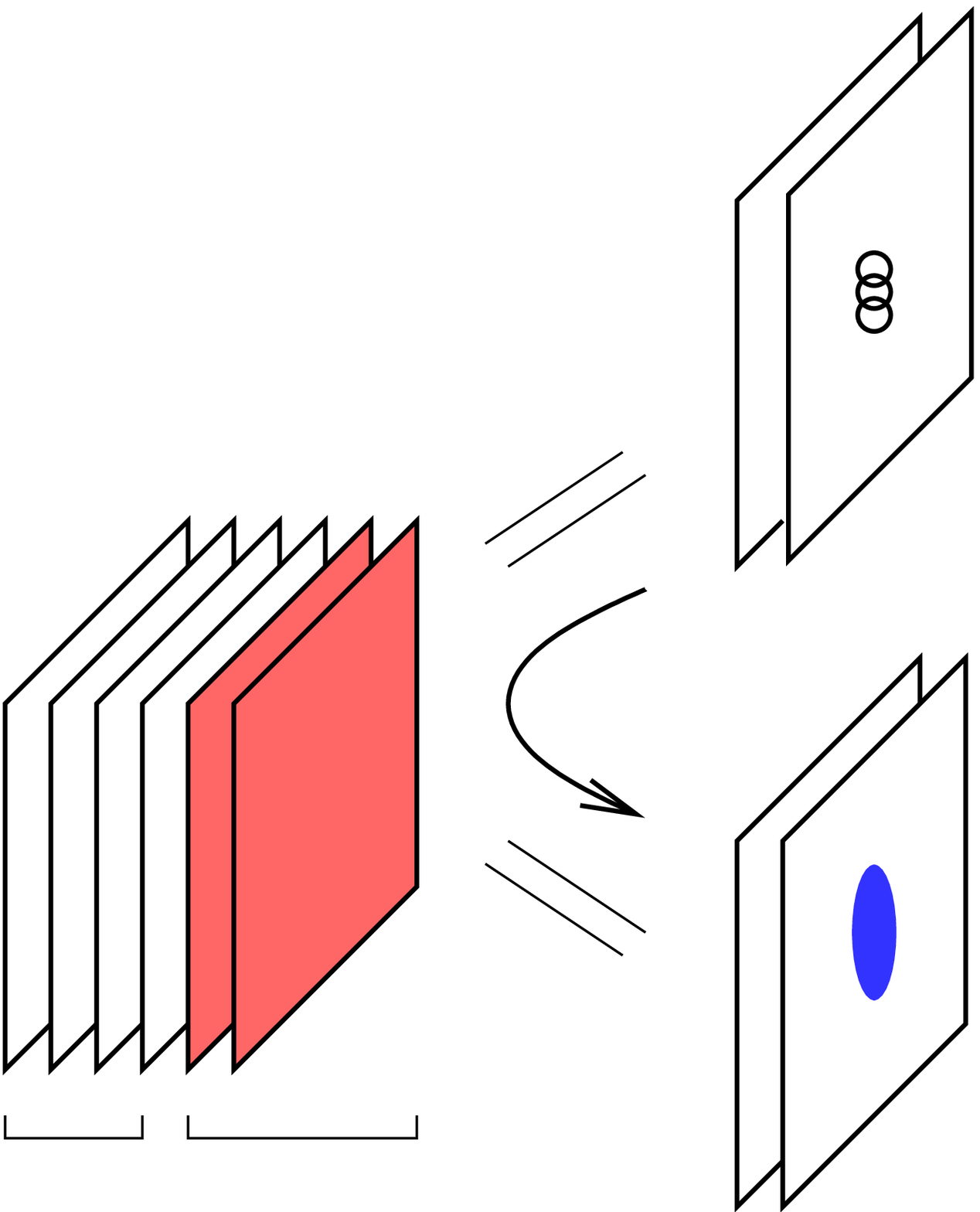}
\put(-100,230){$N$ D4 + $k$ D0}
\put(-35,35){$A_\mu$}
\put(-200,5){$N\!\!+\!2k$}
\put(-190,-8){D4}
\put(-150,2){$2k$ $\overline{\mbox{D4}}$}
\caption{D-brane realization of the ADHM construction via a tachyon
condensation. The ADHM data on the D4-D0 system (upper-right) is
directly translated to the tachyon profile on the D4-anti-D4 system
(left). After the tachyon condensation, it is shown to be equivalent
to just D4-branes with nontrivial gauge field (lower-right).
}
\label{fig1}
  \end{center}
 \end{minipage}
\hspace{5mm}
 \begin{minipage}{7cm}
  \begin{center}
\includegraphics[width=5cm]{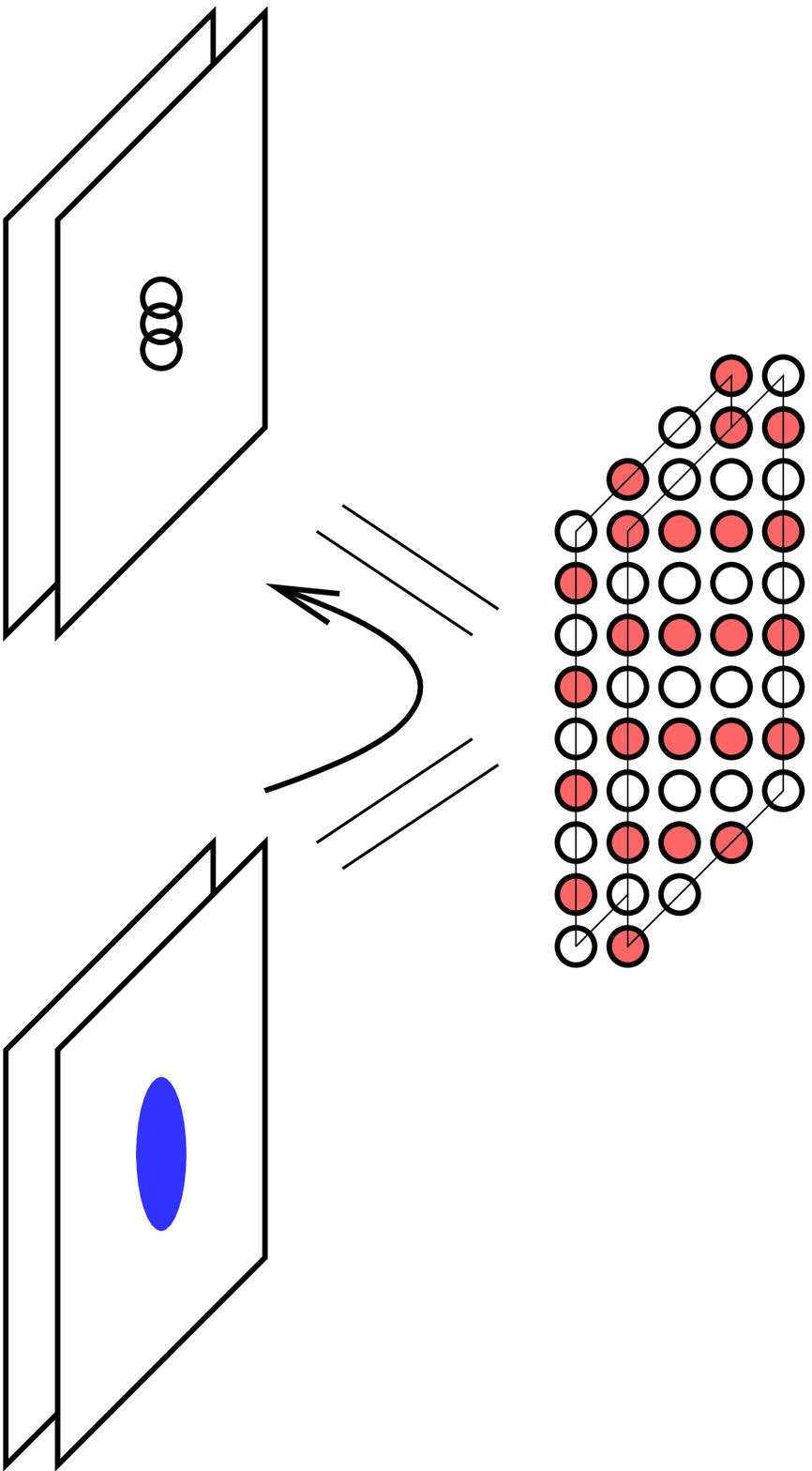}
\put(-85,230){$N$D4 + $k$D0}
\put(-125,30){$A_\mu$}
\put(-55,60){$\infty$D0+$\infty\overline{\mbox{D0}}$}
\caption{The idea of the inverse ADHM construction. The instanton
configuration of the gauge field $A_\mu$ (lower-left) is directly
encoded in the tachyon profile of the $\infty$ number of the
D0-anti-D0-branes (right). In a basis proper for the D0-branes, 
we can read the information of the D0-branes and D0-D4 strings
which are the ADHM data (upper-left).
}
\label{fig2}
  \end{center}
 \end{minipage}
\end{center}
\end{figure}

It turns out that all of the ADHM construction 
and the inverse ADHM construction are
realized in these two ways of understanding of the brane
configurations. In fact, {\it the representation (a) realizes the ADHM
construction, while the representation (b) is nothing but the inverse
ADHM construction.} A schematic picture for the ADHM construction (a) 
is shown in Fig.~\ref{fig1}, and for the inverse ADHM construction (b) 
in Fig.~\ref{fig2}. 

Let us look at the equality (a) more closely. As we shall see in the
next section, the tachyon field, arising from the string connecting the
D4-branes and the anti-D4-branes, has a peculiar form to incorporate the
D0-D0 string excitations ($X^\mu$) and the D0-D4 string excitations
($S$) after the tachyon condensation. Interestingly, an exact treatment
of this leads to the form
\begin{eqnarray}
 T = \lim_{u\to\infty} u \nabla^\dagger \ .
\label{tachyondirac}
\end{eqnarray}
In other words, {\it the Dirac operator is the tachyon}.\footnote{
Relations between tachyons and Dirac operators have been 
discussed in \cite{Hori:1999me,Hellerman:2001bu}.}
The physical essence of the tachyon condensation is that, once the
tachyon expectation value becomes infinite, the corresponding pair of
the D4 and anti-D4-branes disappear. Therefore, from the relation
(\ref{tachyondirac}), the D4-branes surviving
after the tachyon condensation is identified with the zeromodes
of the Dirac operator 
\cite{Terashima:2005ic}.\footnote{See also \cite{Ellwood:2005yz}.} 
$V$ in (\ref{diraczero}) is interpreted as
a ``wave function'' of the remaining D-branes. One can view this
procedure just as a change of basis of the Chan-Paton factor,
and because the basis now depends on $x$, there appears a nontrivial
connection on the remaining D4-branes, which is the gauge field
$A_\mu$ given by the seemingly-unitary transformation of a trivial
connection, (\ref{adhmformula}). In the next section we make this
statement more precise and explain its relation to a Berry's phase on
the worldsheet description of strings in target space background
fields. Fig.~\ref{fig1} shows these processes schematically.

The representation (b) gives in a similar manner the inverse ADHM
construction. Here again, the tachyon profile coming from the strings
connecting the infinite number of pairs of the D0-branes and the
anti-D0-branes is found to be identical to the Dirac operator in
Euclidean 4 dimensions which is a necessary ingredient of the inverse
ADHM construction. See Fig.~\ref{fig2}. Because we need infinite number
of pairs of D0-branes and anti-D0-branes, the tachyon is an infinite
dimensional matrix, which turns out to be a matrix-representation of the
Dirac operator $e^\dagger_\mu(\p_\mu + A_\mu(x))$.

A new outcome of our method using D-branes
is concerning the completeness. Corrigan and Goddard showed explicitly 
\cite{Corrigan:1983sv} that performing the ADHM and the inverse ADHM
constructions succeedingly ends up with going back to the original
configuration, which shows the completeness and the uniqueness.
We find a more direct way of checking the completeness,
without using explicitly the relations (a) and (b): 
in Sec.~\ref{sec4-2} we show that there is a direct
relation between the D4-brane descriptions and the D0-brane description, 
which is horizontal arrow in Fig.~\ref{fig3}.

A surprise is that our ``derivation'' of the ADHM construction 
using the tachyon condensation on the D4-anti-D4 system
turns out to be a
realization of the original derivation of the ADHM construction 
\cite{Atiyah:1978ri}. As well-phrased in Atiyah's lecture note 
\cite{Atiyah:1990ex}, the instanton gauge field in the ADHM construction
is given as an induced 
connection on a subspace of a trivial vector bundle over 
$S^4=P_1({\bf H})$. 
This $P_1({\bf H})$ is a quaternion projective line 
defined by homogeneous coordinates
$(x,y)$ with $x,y\in {\bf H}$ and identified as $(x,y)\sim (xq,yq)$.
The projection onto the sub-bundle is given by a map 
$v(x,y)=Cx+Dy$ which is a $(k+N)\times N$ matrix of quaternions with
constant matrices $C$ and $D$. More precisely, the operator
$\unit_{k+N}-vv^*$ is the projection onto the sub-bundle of our concern.
A certain constraints on $C$ and $D$ ends up with (anti-)self-dual
connections on the sub-bundle, which is the essence of the ADHM
construction. The parameters $C$ and $D$ become the ADHM data 
$S$ and $X^\mu$ after redundant degrees of freedom are gauged away.
In addition, we can choose a gauge $y=1$ in the representation of 
$P_1({\bf H})$. Then, we find that the linear matrix function $v$ is 
realized by our tachyon configuration, and the projection
is given a 
physical interpretation that zeros of the tachyon correspond 
to ``wave functions'' (Chan-Paton factors) of the surviving D-branes. 
The extended space ${\bf H}^{k+N}$ with the trivial bundle 
is nothing but the vector space of the Chan-Paton factor
of the brane-anti-brane system.\footnote{Atiyah's case 
\cite{Atiyah:1990ex} treats ${\bf H}^{N+k}$ resulting in a gauge group
Sp($N$) while in our case the vector space is ${\bf C}^{N+2k}$ for the
gauge group SU$(N)$.} 
The tachyon condensation singles out the sub-bundle with induced
connections on it. 

In our derivation, we haven't referred to any on-shell condition of the
fields appearing and thus to any self-dual equations
(except a condition on the number of the Dirac zeromodes). 
In this sense our
construction works even off-shell.\footnote{At off-shell, often the
boundary state might suffer from divergence and not well-defined, but
our treatment can be justified by the BSFT.} 
The correspondence between
the ADHM data satisfying the ADHM equation (\ref{adhmeq}) 
and the (anti-)self-dual configuration of $A_\mu$ appears once we impose
the supersymmetry condition on both sides.

\section{Derivation of ADHM by D-branes}
\label{sec:3}
\setcounter{footnote}{0}

\subsection{Derivation}
\label{secder}

As briefly described in the previous section, we are interested in
viewing the D0-D4 system solely by D4-branes, by replacing the D0-branes
with pairs of D4-branes and anti-D4-branes accompanied by the tachyon
condensation (the relation (a) in Sec.~\ref{secidea}). 
Eventually this derives the ADHM construction of instantons, as we shall
see. The way we look at the D0-D4 system helps to describe it
rigorously in terms of a boundary state. When D-branes with different
dimensionalities are present, there is a complication in
writing a boundary states of that system because of possible twist
operations (changing boundary conditions) on the boundaries of the string
worldsheet. However if one lifts the D0-branes to the pairs of D4-branes
and anti-D4-branes with the tachyon condensation, 
this complication disappears, which is another motivation for our
description with the tachyons. 

The charges of the $k$ instantons in SU($N$) Yang-Mills theory
is provided by $k$ D0-branes residing on coincident $N$ D4-branes.
To be precise, this correspondence is valid when the gauge field
configuration is at the small instanton singularity. Then the location
of the D0-brane on the D4-brane worldvolume is 
identified with the point-like location of the instantons. 
To obtain instantons with finite size, one has to let the massless mode
of the D0-D4 strings condensate, and roughly speaking, the expectation
value of this massless field on the D0-branes is the size of the
instanton. 

Let us consider first the zero size instantons, equivalently neglecting
the D0-D4 strings. A D0-brane can be described by the following
tachyon condensation on two pairs of a parallel D4-brane and an
anti-D4-brane,  
\begin{eqnarray}
 t=u x^\mu e_\mu^\dagger \ ,
\label{tori}
\end{eqnarray}
with $u\to\infty$ limit.
This is called Atiyah-Bott-Shapiro construction \cite{WittenK,ABS}, 
and in the limit
$u\to\infty$ this configuration becomes a solution of a boundary string
field theory \cite{Kraus:2000nj,Takayanagi:2000rz}, 
and thus is on-shell and a consistent background of string
theory. To have $k$ D0-branes, we prepare $2k$ pairs of
D4-anti-D4-branes.
The location of the D0-branes is encoded as zeros of the tachyon
profile, so to introduce generic location of the $k$ D0-branes, we
generalize (\ref{tori}) to 
\begin{eqnarray}
 t = u (x^\mu \unit_k-X^\mu)\otimes e_\mu^\dagger
\label{t}
\end{eqnarray}
where $X^\mu$ are $k \times k$ constant hermitian matrices.
When $X^\mu$ are simultaneously diagonalizable, it is clear that this
gives the location of the D0-branes after the tachyon condensation. 
Even when they are not, it has been shown that this incorporation 
(\ref{t}) of the 
$X^\mu$ matrices with the D4-anti-D4-brane boundary state results in
a D0-brane boundary state with transverse scalar field profile
$X^\mu$ \cite{Asakawa:2002ui}, thus (\ref{t}) is the correct profile
including the massless excitation of the D0-D0 strings.

The total system of our concern consists of 
$N+2k$ D4-branes and $2k$ anti D4-branes, thus the tachyon $T$
in the system
is a complex $2k\times (N+2k)$ matrix. The system has the
gauge invariance U($N+2k$)$\times$U($2k$), and the tachyon is in a
bi-fundamental representation with respect to this gauge symmetry.
The low-lying excitations of the strings also include the gauge fields
on the D4-branes and the anti-D4-branes, $A_\mu^{{\rm D4}}(x)$ 
and $A_\mu^{{\rm antiD4}}(x)$. We put them vanishing, 
$A_\mu^{{\rm D4}}(x)= A_\mu^{{\rm antiD4}}(x)=0$. 
These low-lying excitations of the D4-anti-D4-branes can be conveniently
written as an $(N+4k)\times(N+4k)$ matrix, 
\begin{eqnarray}
M= 
\bordermatrix{
& \overbrace{\hspace{10mm}}^{N+2k} 
& \overbrace{\hspace{15mm}}^{2k} \cr\nonumber\\[-8mm]
& A^{\rm D4} & T^\dagger \cr
& T & A^{\rm antiD4}
\cr}
\begin{array}{l}
\bigm\}
{\scriptstyle N+2k}
 \\
\bigm\}
{\scriptstyle 2k}
\end{array}
\ ,
\label{M}
\end{eqnarray}
which is known as a superconnection.
Then the gauge symmetry U($N+2k$)$\times$U($2k$) acts as
$M \rightarrow U^\dagger M U+U^\dagger d U,$ 
$U={\rm diag} (U_1,U_2)$ where $U_1 \in $U($N+2k$), $U_2 \in $U($2k$).
For the present case, the previous tachyon $t$ (\ref{t}) is embedded
in the tachyon $T$ as 
\begin{eqnarray}
T=
\bordermatrix{
& \overbrace{\hspace{10mm}}^{N} 
& \overbrace{\hspace{15mm}}^{2k} \cr\nonumber\\[-8mm]
& 0 & t
\cr}
\Bigm\}
{\scriptstyle 2k}
\ .
\end{eqnarray}
where the entry ``0'' means a vanishing matrix of the
size $2k \times N$. 

In general, nothing prevents us from turning on this vanishing part 
of the tachyon matrix $T$. In fact, this part should correspond 
to the excitation of the string connecting the remaining $N$ D4-branes
and the created $k$ D0-branes. This is obvious when we look at the
matrix (\ref{M}). The lower-right $4k\times 4k$ corner becomes the 
$k$ D0-branes after the tachyon condensation $u\to\infty$, so the
lower-left corner should represent the D0-D4 strings.
Let us turn on generic value in the left half entries of the
tachyon matrix $T$ as 
\begin{eqnarray}
 T = \lim_{u\to\infty}
\Bigl( \hspace{5mm} u S^\dagger \hspace{5mm} t\hspace{5mm}
\Bigr) \ .
\label{genericT}
\end{eqnarray}
where $S$ is a constant complex $N\times 2k$ matrix.\footnote{
We expect $X^\mu$ dependent parts of $S$ corresponds 
to the massive excitations.
} 
The indices which $S$ carries in fact coincide with that of the massless
excitation of the strings connecting $k$ D0-branes and $N$ D4-branes:
it is known that a field of fundamental representation in U($k$)
appears from the D0-D4 strings, 
and it is charged under the ``global'' U$(N)$ as a fundamental
and in the 
(${\bf 2}, {\bf 1}$) and (${\bf 1}, {\bf 2}$) representations of 
the ``internal'' rotation group SO(4)$\sim$SU(2)$\times$SU(2) 
acting on the worldvolume of the D4-branes. At this stage, we
established the equality of the upper-right and the left figures in
Fig.~\ref{fig1}. 

Then with this nonzero $S$, which D-branes remain after the 
tachyon condensation? The answer to this question is another equality 
connecting the left and the lower-right figures in Fig.~\ref{fig1}. 
In the BSFT,
the pair of D4 and anti-D4-branes with infinite value of the tachyon
vanishes, while if the tachyon remains zero those D-branes survive. 
Now the tachyon is a $2k\times(N+2k)$ matrix, so one has to diagonalize
the whole tachyon matrix by a gauge transformation of 
U($N+2k$)$\times$U($2k$),
\begin{eqnarray}
 T \to T'=U_2^\dagger T U_1
\label{t'}
\end{eqnarray}
where $U_i$ is the gauge transformation associated with the gauge field
$A^{(i)}$. We can use this gauge degrees of freedom to get the following 
canonical form of the tachyon, 
\begin{eqnarray}
T'=\lim_{u\to\infty} u
\bordermatrix{
& \overbrace{\hspace{10mm}}^{N} 
& \overbrace{\hspace{15mm}}^{2k} \cr\nonumber\\[-8mm]
& 0 & 
\begin{array}{ccc}
 * & 0 & 0 \\
 0 & * & 0 \\
 0 & 0 & *
\end{array}
\cr}
\Biggm\}
{\scriptstyle 2k}
\label{t''}
\end{eqnarray}
where the left half of the matrix is vanishing while the right half 
($2k\times 2k$) is diagonal with nonzero entries.\footnote{This
nondegeneracy condition is an assumption of the ADHM construction.}
Generically this form of the matrix is available.

In this rotated basis of the Chan-Paton factor, 
it is easy to figure out which brane is surviving in 
the $u\to\infty$ limit of the tachyon condensation. 
The D4-branes corresponding to the left half (column $1, \cdots,N$) 
are surviving the annihilation process while the right half
(column $N+1,\cdots,N+2k$) will be pair-annihilated with the $2k$
anti-D4-branes.

Let us look at the properties of the remaining $N$ D4-branes. 
Now according to the above gauge transformation $U_1$ and $U_2$, 
we have actually a nonzero gauge field on the $(N+2k)$ D4-branes,
\begin{eqnarray}
 A_\mu^{(1)} = U_1^\dagger \p_\mu U_1 \ .
\end{eqnarray}
But what we need is only a part of this matrix, 
given by the $ij$ components ($i,j = 1,\cdots,N$), 
because other Chan-Paton indices are for disappearing D4-branes and
unphysical.  
For the physical gauge field on the remaining $N$ D4-branes,
we need only a part of the information of the gauge rotation matrix
$U_1$. If we explicitly write the matrix $U_1$ as 
\begin{eqnarray}
 U_1 = 
\bordermatrix{
& \overbrace{\hspace{10mm}}^{N} 
& \overbrace{\hspace{15mm}}^{2k} \cr\nonumber\\[-8mm]
& V & V'
\cr}
\Bigm\}
{\scriptstyle N+2k}
\ ,
\end{eqnarray}
then our physical part ($N\times N$) of the gauge field is
given by 
\begin{eqnarray}
 \left[A_\mu^{(1)}\right]_{N\times N} = V^\dagger \p_\mu V \ .
\label{adhmformulader}
\end{eqnarray}
This (\ref{adhmformulader}) is nothing but the ADHM formula
(\ref{adhmformula}).
We can conclude that we deduce
the ADHM construction
if this $V$ is a collection of the normalized zeromodes of the Dirac
operator (\ref{diraczero}).
And this is in fact the case. First, the normalization condition
required in the ADHM construction, $V^\dagger V = \unit_N$,
is just a part of the unitarity condition of $U_1$. So this is
satisfied. Second, the zeromode condition (\ref{diraczero})
is found to be just a part of the unitary rotation (\ref{t'}),
because the Dirac operator is exactly the tachyon field $T$, and 
the rotated form of $T'$ has zeros as in (\ref{t''}). 

Here completes the derivation of the ADHM construction from a tachyon
condensation in D-branes. We provide a rigorous proof in
Sec.~\ref{secb} by realizing the tachyon condensation procedure in the
boundary state formalism. We will see that this relation between the
data $X^\mu, S$ on the unstable D4-anti-D4-branes and the gauge
fields $A_\mu$ after the tachyon condensation is exact, and thus those
two descriptions are equivalent.

We used a unitary transformation for the rotation of the basis of
the Chan-Paton factor, but this is equivalently described by just 
the notion of the zeromode eigen function of the tachyon matrix $T$,
as originally described in \cite{Terashima:2005ic}. In this terminology, 
the gauge field $A_\mu$ is provided as a Berry's phase as in
our previous paper for the Nahm construction \cite{Hashimoto:2005yy}: 
the zeromode eigen states 
($V_i(x) = \langle x|V_i\rangle$, $i=1,\cdots,N$) are functions of $x$,
and furthermore, $x$ is a function of the worldsheet boundary time
$\sigma$, therefore on the worldsheet action, the Berry's phase is
induced,
\begin{eqnarray}
\gamma_{ij} = 
\oint \! d\sigma 
\left\langle V_i \bigl|\p_\sigma
\bigr|V_j
\right\rangle
=
\oint \! d\sigma 
\p_\sigma X^\mu(\sigma)
\left\langle V_i \bigl|\p_\mu
\bigr|V_j
\right\rangle \ .
\end{eqnarray}
This is a worldsheet boundary coupling to a background gauge field
given by the coefficients,
\begin{eqnarray}
 [A_\mu]_{ij}= \left\langle V_i \bigl|\p_\mu
\bigr|V_j\right\rangle \ ,
\end{eqnarray}
which is the ADHM formula. 

\subsection{Noncommutative ADHM and identification of $S$}
\label{secnc}

One of the recent interesting topics has been solitons on noncommutative 
spaces, which was initiated by Nekrasov and Schwarz 
\cite{Nekrasov:1998ss} who related the 
resolution of the small instanton singularity in the ADHM moduli space
with the noncommutativity. In \cite{Nekrasov:1998ss}, how the
ADHM construction in the noncommutative space works was explained: the
noncommutative ADHM construction is obtained simply by replacing all the
procedures in the ADHM construction by their noncommutative
generalization. The product is replaced with Moyal $*$ product, and the
ADHM equation is modified to have a resolution of the singularity.
In this subsection we derive the noncommutative ADHM construction, and 
explain why this works in this way, in terms of D-branes and the
tachyon condensation. 

As seen also in \cite{Nekrasov:1998ss}, the space
noncommutativity is introduced as a background constant NS-NS $B$-field
on the worldvolume of the D-branes \cite{Connes:1997cr,Seiberg:1999vs}.  
So let us think of putting all the brane setup in the background
constant $B$-field. 
The background $B$-field effectively induces a constant field strength
on the D-branes, $F_{\mu\nu}=B_{\mu\nu}/2\pi\alpha'$. 
In the language of the boundary
state of the D-branes, this simply induces a term 
$\oint d\sigma F_{\mu\nu}x^\mu(\sigma)\dot{x}^\nu(\sigma)$ in the 
boundary action, and nothing more than that. 
We have to perform the Seiberg-Witten map \cite{Seiberg:1999vs} to 
obtain the
description in terms of fields in the equivalent noncommutative space.
This makes things complicated, and furthermore for finite $\alpha'$
there is no known explicit expression for the Seiberg-Witten map.
That is to say, the elegant noncommutative ADHM construction,
with just the noncommutative zeromode equations
(\ref{diraczero}) and the noncommutative overlap 
(\ref{adhmformula}), appear to be difficult to show up in this attempt. 

Instead of this trivial trial, we take a different route to realize
the noncommutativity, which turns out to lead us to the realization of
the noncommutative ADHM construction. Consider a single
D4-brane. Putting it in the constant $B$-field is equivalent on the 
worldvolume to regard the D4-brane as a bound state of infinitely many
D0-branes \cite{Ishibashi:1998ni,Seiberg:2000zk}. Note that this is
possible without the $\alpha'\to 0$ limit \cite{Ishibashi:1998ni}.
This is a famous example of Matrix theory. The transverse
scalars of the D0-branes are turned on as $\Phi_\mu = \hat{x}_\mu$
($\mu=1,2,3,4$)
where infinite dimensional matrices $\hat{x}_\mu$ satisfy the 
noncommutative algebra\footnote{Note that this $\hat{x}$ is different
from that appearing in the usual quantum mechanics where 
the Heisenberg algebra $[\hat{x}_\mu, \hat{p}_\nu]=i\delta_{\mu,\nu}$ 
is satisfied. The latter will be used in Sec.~\ref{sec:4}.}
\begin{eqnarray}
 [\hat{x}_1, \hat{x}_2] = i\theta_{12}=-i\alpha'/B_{12} \ , \quad
 [\hat{x}_3, \hat{x}_4] = i\theta_{34}=-i\alpha'/B_{34} \ .
\label{nc}
\end{eqnarray}
This is the appearance of the noncommutativity. Here $\hat{x}_\mu$ 
are infinite dimensional matrices, and their explicit expression is
given by 
\begin{eqnarray}
&&[\hat{x}_1 + i \hat{x}_2]_{(n_1,n_2),(m_1,m_2)} = \sqrt{2\theta_{12}}
\sqrt{n_1}\delta_{n_1,n_2-1}\delta_{m_1,m_2} \ , 
\nonumber \\
&&
[\hat{x}_3 + i \hat{x}_4]_{(n_1,n_2),(m_1,m_2)} = \sqrt{2\theta_{34}}
\sqrt{m_1}\delta_{n_1,n_2}\delta_{m_1,m_2-1} \ .
\end{eqnarray}
Henceforth, we use this matrix representation for the $N+2k$ D4-branes
and $2k$ anti-D4-branes. We will find that this way of considering the
noncommutativity leads to the noncommutative ADHM construction.
Note that for the anti-D4-branes, we consider the same
transverse scalar field configuration (\ref{nc}) of anti-D0-branes. 

As a warm-up, we consider the example of a pair of 2 D4-branes and 2
anti-D4-branes. We know that the tachyon configuration on these pairs
(\ref{tori}) produces a single D0-brane after the tachyon condensation 
$u\to\infty$. What about the case with the noncommutativities? 
The D4-branes consist of infinite number of D0-branes, while the
anti-D4-branes are made of anti-D0-branes.\footnote{
The total charge of the
D0-branes is vanishing, as easily seen in the Ramond-Ramond coupling in
the BSFT action of the brane-anti-brane
\cite{Kraus:2000nj,Takayanagi:2000rz} which reads $\int C_0\wedge
(e^B-e^B) = 0.$} 
Because we have the D0-branes and the anti-D0-branes, there exists
a complex tachyon field as an excitation of a string connecting those.
It turns out that the tachyon profile (\ref{tori}) with replacement of
$x_\mu$ with the matrix $\hat{x}_\mu$, 
\begin{eqnarray}
 t=u (\hat{x}^\mu-X_\mu\unit_\infty) e_\mu^\dagger 
\label{t=unc}
\end{eqnarray}
with the limit $u\to\infty$ is a solution of a BSFT. 
Here $X_\mu$ are constant parameters.
To see this, we apply the idea of
\cite{Terashima:2005ic,Ellwood:2005yz} for the above tachyon profile
(\ref{t=unc}).
The zeromode of the above matrix specifies the remaining D0-brane. 
In fact, there exists a single zeromode given by
\begin{eqnarray}
&&\Psi= \exp\left[
-\frac{X_1^2+X_2^2}{4\theta_{12}}-\frac{X_3^2+X_4^2}{4\theta_{34}}
\right]\left(
\begin{array}{c}
1 \\0
\end{array}
\right)
\nonumber \\
&&\hspace{10mm} 
\otimes \left[
\sum_{n=0}^\infty \frac{1}{\sqrt{n!}}\left(
\frac{X_1+iX_2}{\sqrt{2\theta_{12}}}
\right)^n
\right]
\left[
\sum_{m=0}^\infty \frac{1}{\sqrt{m!}}\left(
\frac{X_3+iX_4}{\sqrt{2\theta_{34}}}
\right)^m\right]\ket{n,m}
\nonumber
\end{eqnarray}
where the first 2-vector is for the vector space on which 
$e_\mu^\dagger$ acts,
and we have chosen the representation of the base vector space of the
noncommutative operators as the standard one labelled by 
$\ket{n,m}$ $(n,m=0,1,2,\cdots)$ (we need a tensor
product of two Hilbert spaces since we are working in $4=2+2$ 
dimensions). We included the normalization factor already in $\Psi$. 

The location of the surviving D0-brane can be found by the vacuum
expectation value of $\hat{x}_\mu$ or in other words, the scalar field 
matrix element with the index given by the above $\Psi$, as
\begin{eqnarray}
 \Psi^\dagger \hat{x}_\mu\Psi = X_\mu \ ,
\label{locsurd0}
\end{eqnarray}
as anticipated.

The generalization of the tachyon profile
(\ref{tori}) to the noncommutative case is given by (\ref{t=unc}),
therefore, the noncommutative generalization of the full tachyon
operator (\ref{genericT}) concerning the ADHM construction 
should be provided by replacing $x_\mu$ with by the
infinite dimensional matrix $\hat{x}_\mu$. The computation of finding 
zeromodes can be done in the infinite dimensional matrix
multiplications, and this is nothing but working with Moyal $*$-product
with usual $x_\mu$. Thus we have derived the noncommutative ADHM
construction.  

In the previous subsection, we have identified a part of the tachyon
matrix $S$ as an excitation of the D0-D4 strings. There we presented
an argument that this $S$ carries a correct charge of the strings.
Here we show that, for small fluctuation of $S$, this gives 
the mass spectrum identical with the fluctuation of the D0-D4 strings.
An explicit instanton configuration in U$(2)$ noncommutative Yang-Mills
theory was given in \cite{Furuuchi:2000dx} via the noncommutative
ADHM construction \cite{Nekrasov:1998ss}. There explicit construction
with the parameter $S$ results in the following instanton configuration
of the gauge field (see Eqs.~(5.10)--(5.12) of
\cite{Hashimoto:2001pc} where $S$ is written as $\rho$):
\begin{eqnarray}
 \hat{x}_\mu-\theta_{\mu\nu}A_\nu
= U_0^\dagger \hat{x}_\mu U_0
+ 
\biggl[
\hat{x}_\mu \ket{0,0}\bra{0,0}+ \ket{0,0}\bra{0,0}\hat{x}_\mu 
\biggr]\otimes
\left(
\begin{array}{cc}
0 & 0 \\ 0& \rho
\end{array}
\right)
+ {\cal O}(\rho^2) \ ,
\nonumber
\end{eqnarray}
where $U_0$ is a shift operator which shifts the Hilbert space index by
one, $U_0\ket{s}=\ket{s+1}$ where 
$\ket{s=(n+m)(n+m+1)/2 +m}\equiv\ket{n,m}$. The first term in the right
hand side of the above solution is the well-known noncommutative soliton
generated by the shift operator \cite{Gopakumar:2000zd}. The rest terms
are a deviation from the shift-operator-generated noncommutative
soliton, and they come in the first off-diagonal entries in this Hilbert
space, as specified by 
$\hat{x}_\mu \ket{0,0}\bra{0,0}$ or $\ket{0,0}\bra{0,0}\hat{x}_\mu$.
These entries are nothing but the ones giving the mass spectrum of the
hypermultiplets coming from the D0-D4 strings, as shown in
\cite{Aganagic:2000mh}. 
Therefore in this noncommutative example, through the ADHM construction,
it is explicitly shown that the matrix $S$ appearing in a part of the
tachyon is in fact the D0-D4 string excitation.
This also implies that 
the normalization of $S$ in (\ref{genericT}) is indeed correct. 

\subsection{Exactness shown in boundary state formalism}
\label{secb}

The derivation of the ADHM construction in terms of D-branes
presented in Sec.~\ref{secder} is just an analysis of the bases of the
matrix-valued tachyon field. Nevertheless, the equivalence of the 
D0-D4 system and the D4-branes with instanton gauge fields can hold
beyond the $\alpha'$ corrections, which we will show in this subsection.
We show this by using a boundary state
formalism intimately related to the BSFT.
Boundary states are ``states'' in the closed string Hilbert space,
specified by boundary conditions on the string worldsheet. 
The quantized world sheet scalar fields in the closed string picture, 
${\widehat X}^\mu (\sigma)$, act on them.
The boundary state is one of the definitions of D-branes,
thus once one can prove that two boundary states are equal, it
immediately shows that those two D-branes are identical.

The action of a BSFT for brane-anti-branes was
constructed in \cite{Kraus:2000nj,Takayanagi:2000rz} with worldsheet 
boundary interactions including tachyons. This is straightforwardly 
generalized to boundary states, whose useful expression can be found in
K-matrix theory, \cite{Asakawa:2002ui}.\footnote{
However, all of the situations considered in the literature 
(except \cite{Jones:2003ae}) have dealt with equal number of
D-branes and anti-D-branes which makes it possible to trade the
(gamma) matrices appearing in the boundary interaction for
boundary fermions. But in our present case, since the number of
D4-branes is different from that of the anti-D4-branes, we cannot use
the fermion representation. 
Instead, we use the explicit matrix formula for the boundary
interaction.} For simplicity, 
we ignore all the worldsheet fermions and ghosts which 
are not relevant for our purpose.
Then
the boundary state is given as
\begin{eqnarray}
 \ket{B}= \int [dx] \; e^{-S_{\rm b}}\ket{x} \ .
\end{eqnarray}
The ket $\ket{x}$ is an eigenstate of the closed string worldsheet 
scalar coordinates
${\widehat X}^\mu$, ${\widehat X}^\mu \ket{x}=x^\mu(\sigma)\ket{x}$,
where $\sigma$ 
parameterizes the boundary of the string worldsheet.
The boundary perturbation $e^{-S_{\rm b}}$
is represented as a partition function
for a quantum mechanics with Hamiltonian $(M_0)^2$ 
acting on a finite ($N+4k$) dimensional Hilbert space:
\begin{eqnarray}
e^{-S_{\rm b}}={\rm Tr}_{(N+4k) \times (N+4k)} {\rm P} 
\exp\left[-\int\! d\sigma (M_0)^2\right], 
\quad
M_0 
\equiv
\left(
\begin{array}{cc}
0 & T(x)^\dagger
\\
T(x) &
0
\end{array}
\right) \ .
\label{bp}
\end{eqnarray}
We may substitute the tachyon configuration (\ref{genericT}) 
to represent the
system of D0-D4-branes.
Note that the system is finite in the sense that the Hilbert space is
finite dimensional. This is in contrast to the situation we found in
the derivation of the Nahm construction of monopoles
\cite{Hashimoto:2005yy}. 

Here $x^\mu$ in $T(x)$ is considered as the world sheet string
coordinate $x^\mu(\sigma)$. This $x^\mu(\sigma)$ has another important 
interpretation: 
a time dependent external field in the quantum mechanics governed by 
the Hamiltonian $(M_0)^2$, where $\sigma$ is the (Euclidean) 
time of the quantum
mechanical system. With this in mind, 
let us consider the tachyon condensation
$u\to\infty$. We can diagonalize $M_0$ as in Sec.~\ref{secder} by the
matrices $U_i$, which depend on $x(\sigma)$. In the $u \rightarrow
\infty$ limit, we find that the remaining terms in 
the path-ordered trace is just the
$N\times N$ part, because the diagonalized $M_0$ has only $N$ vanishing
eigenvalues. The other non-zero eigenvalues give a vanishing trace due
to the limit $u\to\infty$. In this ``selection'' of the $N$ eigenmodes
in the quantum mechanical system, note that the transformation $U_i$
depend on $\sigma$ through $x(\sigma)$. In other words, 
the wave function of the D-brane in the Chan-Paton space is a function
of $x$ and thus of $\sigma$. Therefore  
a Berry's phase $U_i^\dagger \p U_i$ should be associated with it. 
This phase is exact because the $u \rightarrow \infty$ limit is
the same as the adiabatic limit in the quantum mechanics interpretation
(see \cite{Hashimoto:2005yy} and
also \cite{Asakawa:2002ui}, where the kinetic term for $x(\sigma)$
disappears in the $u\to\infty$ limit). Thus we exactly have
\begin{eqnarray}
e^{-S_b}={\rm Tr}_{N \times N} {\rm P} 
\exp\left[-\int d \sigma A_\mu(x) \p_\sigma x^\mu(\sigma)\right]\ .
\label{bp2}
\end{eqnarray}
This is the boundary perturbation for 
the $N$ D4-branes with the gauge field $A_\mu$, on the boundary state.

We have shown that the boundary state of the D0-D4 system (\ref{bp})
is identical with the boundary state of the D4-branes with instanton
gauge field (\ref{bp2}). This proves that the procedures of the
tachyon condensation in Sec.~\ref{secder} is valid in string theory.
An important point is that this also gives a
strong evidence that the instanton configurations on the D4-branes do
not receive $\alpha'$ corrections. Basically there have been no reason
to believe that small instantons, where the scale of the instanton gets
small and the curvature is not slowly varying, do not receive any 
stringy corrections of $\alpha'$. But here we have shown that the
small instantons singularity limit of the self-dual configuration
corresponding to vanishing $S$  
is string-theoretically {\it equivalent} to the D0-brane description
and thus provides a worldsheet conformal point.

\section{Inverse ADHM and Completeness from D-branes}
\label{sec:4}
\setcounter{footnote}{0}

The inverse ADHM construction is somewhat mysterious from the view point
of obtaining induced connections on a sub-manifold. However the benefit
of considering the ADHM construction in terms of D-branes is that also 
this inverse procedure is easily derived, 
owing to the democratic nature of 
D-branes. Instead of using the D-brane descent relations for the
tachyon condensation, here we use the D-brane ascent relation found in
\cite{Terashima:2001jc} and developed in \cite{kmatrix,Asakawa:2002ui}.
As we shall see, it realizes the inverse ADHM construction, and the
philosophy is depicted in Fig.~\ref{fig2}. 

The power of the ADHM / inverse ADHM constructions is the uniqueness and
the completeness. For each instanton solutions there is a corresponding
ADHM data, and vice versa. This was explicitly shown
\cite{Corrigan:1983sv} by applying the ADHM 
and the inverse ADHM procedure succeedingly. Our D-brane realization 
enables one to
access this completeness much more easily: 
we can show directly that D0-brane
configurations used in the ADHM and the inverse ADHM are the same,
which gives a direct proof of the completeness without referring
explicitly to the procedures of the ADHM / inverse ADHM constructions.

\subsection{Derivation of inverse ADHM construction}
\label{secderin}

We start with giving a brief summary of the inverse ADHM construction
for a reference. The inverse ADHM construction is a way to get the
original ADHM data $(X^\mu, S)$ from a given instanton configuration
$A_\mu(x)$.  One starts with computing normalized Dirac zeromodes,
\begin{eqnarray}
D^\dagger \psi = 0 \ , \quad D \equiv e_\mu D_\mu \ , 
\quad 
D_\mu \equiv \p_\mu + A_\mu \ . 
\label{zeromodeD}
\end{eqnarray}
Note that $D^\dagger = e_\mu^\dagger D_\mu$ 
has two spinor indices and accordingly $\psi$ has a spinor index
$\alpha=1,2$ which we often omit. Then Atiyah-Singer index theorem
\cite{Atiyah:1968mp} ensures that there are $k$ 
normalizable zeromodes $\psi_i(x)$ labeled by $i=1,2,\cdots,k$ 
satisfying 
\begin{eqnarray}
 \int\! d^4x\; \psi^\dagger_i \psi_j = [\unit_{k}]_{ij} \ .
\label{normalpi}
\end{eqnarray}
Usually the Dirac operator is defined as $\gamma_\mu
D_\mu$ where $\gamma_\mu = \left(\begin{array}{cc} 0 & e_\mu \\
e_\mu^\dagger & 0\end{array}\right)$, but here we call the chiral
decomposed operator as a Dirac operator. The zeromode $\psi$ has
negative chirality, while there is no normalizable zeromode of
$D$ (which has positive chirality).

Let us consider some non-normalizable scalar zeromodes of the Laplacian
$D_\mu D_\mu$ (note that for self-dual gauge fields we have a relation
$e_\mu^\dagger D_\mu e_\nu D_\nu = D_\mu D_\mu\otimes\unit_2$), 
which is going
to be another important ingredient in the inverse ADHM construction. 
There are
$N$ non-normalizable zeromodes $\phi_a(x) (a=1,2,\cdots,N)$. 
When there is no instanton background, 
these reduce to constant wave functions. 
The normalization of $\phi$ is determined in such a way that 
in the asymptotic region $x^2 \gg 1$
they coincide with the original constant wave functions up to a
certain SU$(N)$ gauge transformation. Since 
the instantons are localized near the origin,
in the asymptotic region the instanton gauge field should be written
as a pure gauge $A_\mu \sim g^\dagger \p_\mu g$.
Thus, if we align the $N$ zeromodes to form an $N\times N$ matrix,
it coincides with $g^\dagger$ times the original constant wave
functions.

Using these spinors and scalar zeromodes, 
one can reconstruct the ADHM data by the 
following formulas, 
\begin{eqnarray}
&& [X_\mu]_{ij} = \int\! d^4x \; \psi_i^\dagger x_\mu \psi_j \ , 
\label{inverseADHM1}
\\
&& \left[S\right]_{ia\alpha} 
= \frac{1}{2\pi}\int \! d^4x \; [\psi_i^\dagger e_\mu]_\alpha D_\mu
\phi_a \ .
\label{inverseADHM2}
\end{eqnarray}
The second relation is not the familiar one written in 
\cite{Corrigan:1983sv} but this is the original one which can be 
found for example in \cite{Nahmtalk}. This expression turns out to be
closely related to our D-brane derivation.

Stringy derivation of this inverse ADHM construction is just the
realization of the D0-D4 system in terms of infinite number of D0-branes 
and anti-D0-branes (see Fig.~\ref{fig2}).
According to the BSFT, D4-branes with nontrivial
gauge fields on them are realized by a tachyon condensation of 
infinite number of pairs of D0-branes and anti-D0-branes. 
The precise and exact field profiles on those D0-branes are
the following tachyon condensation and the transverse scalar 
field \cite{Terashima:2001jc,kmatrix,Asakawa:2002ui}\footnote{
One may write this set as a superconnection in which the gauge
transformation property is easy to read,
\begin{eqnarray}
 \left(
\begin{array}{cc}
 \hat{x}_i & T^\dagger \\
T & \hat{x}_i
\end{array}
\right) \ .
\label{xttx}
\end{eqnarray}
}
\begin{eqnarray}
&& T 
= \lim_{u\to\infty} u
 (\hat{p}_\mu\otimes \unit_{N} - iA_\mu(\hat{x}))\otimes e_\mu^\dagger
= -i \lim_{u\to\infty} u D^\dagger \ , \nonumber \\
&&
\Phi_\mu^{\rm D0} = \Phi_\mu^{\rm antiD0} = 
\hat{x}_\mu\otimes\unit_{N}\otimes \unit_2 \ .
\label{tdirac}
\end{eqnarray}
Note that $\hat{p}$ and $\hat{x}$ are infinite dimensional matrix
representation of the Heisenberg algebra, $[\hat{x},\hat{p}]=i$,
and thus the Dirac operator is in an infinite dimensional matrix
representation. 
In the limit $u\to\infty$ one can show via the boundary state formalism
that this tachyon configuration is exactly equivalent to the D4-brane
configuration with the gauge field $A_\mu(x)$. 

At this stage, it is already clear that the tachyon is in fact the Dirac
operator in the inverse ADHM construction.
We apply the philosophy of the tachyon condensation in which only the
Chan-Paton indices with zero 
tachyon eigenvalues survive in the $u \rightarrow \infty$ limit 
\cite{Terashima:2005ic}. 
Then what is
important is the explicit zeromodes of the tachyon $T$, 
\begin{eqnarray}
\biggl[
  (\hat{p}_\mu\otimes \unit_{N} - iA_\mu(\hat{x}))\otimes e_\mu^\dagger
\biggr] 
|\psi\rangle =0 \ .
\end{eqnarray}
Note that the zeomodes of $T$ correspond to D0-branes
and those of $T^\dagger$ correspond to anti-D0-branes.
As usual in quantum mechanics, inserting a complete set
\begin{eqnarray}
 \int \! d^4x |x\rangle\langle x| = \unit_{\infty}
\label{completeset}
\end{eqnarray}
where $|x\rangle=|x_1,x_2,x_3,x_4\rangle$ 
is the eigen vector of the matrix $\hat{x}$,
\begin{eqnarray}
 \hat{x}_\mu |x_1, x_2, x_3, x_4\rangle = 
 x_\mu |x_1, x_2, x_3, x_4\rangle \ , 
\end{eqnarray}
we recover the relation (\ref{zeromodeD}) with the definition 
$\psi(x) = \langle x | \psi \rangle$. 
(In this section, for notational simplicity, 
we sometimes omit the spinor and the U$(N)$ indices, which
are not relevant below.)
Furthermore, the
normalization of the infinite dimensional vector $|\psi\rangle$ is given
by $\langle \psi | \psi \rangle=1$ which is, by again inserting the
complete set (\ref{completeset}), shown to be equivalent to 
(\ref{normalpi}). 

On the off-shell boundary states (or the BSFT), 
the worldsheet boundary interaction appears in the form (\ref{bp}),
and thus in effect the tachyon always appears as a combination
$TT^\dagger$ or $T^\dagger T$. So the important is the 
zeromodes (with positive chirality)
$|\phi\rangle$
of $TT^\dagger =u^2 D_\mu D_\mu\otimes \unit_2$ in this sense. 
(There may be 
non-normalizable zeromodes of $T^\dagger T$ with negative chirality.) 
Since the Dirac operator
is written by an infinite dimensional matrix, the non-normalizable 
zeromode $|\phi\rangle$ is an infinite dimensional vector.  
In the
absence of the gauge fields, this non-normalizable zeromode is just 
a vector state $|p=0\rangle$ (with the spinor and gauge indices) 
in the expression of the momentum
eigenstates. The reason is that when $A_\mu=0$, the BSFT 
tachyon potential is just $e^{-u^2p^2}$ which,
in the limit $u\to\infty$, removes all the momentum states except the
zeromodes.\footnote{
More precisely, since the Dirac operator has non-normalizable zeromodes
and then has a continuous spectrum near the zeromodes,
we should keep non-zeromodes which are very close to the zeromodes.
This is because the D4-branes can not be described by D0-branes only,
in this situation.}
This $|p=0\rangle$ 
corresponds to a D4-brane. Here the normalization was fixed as usual.
In the $x$ representation, these non-normalizable zeromodes are just
constant.
When instanton gauge fields are turned on, 
we will have $2N$ non-normalizable zeromodes of $T T^\dagger$
with positive chirality, 
$\ket{\phi}_{a,j}$ where $a=1,2,\cdots,N$ and $j=1,2$. The index $a$
is for the SU($N$) gauge group, and $j$ is trivially related to
the spinor index because $TT^\dagger$ is proportional to $\unit_2$. 
Using the scalar $\phi_a(x)$, these are written as 
$\langle x|\phi\rangle_{a,j}= \phi_a(x)\otimes c_j$ where
$c_j$ are  constant spinors,
$c_1 = $ {\scriptsize$\left(\begin{array}{c}1 \\  0\end{array}\right)$} 
and  
$c_2 = $ {\scriptsize$\left(\begin{array}{c}0 \\ 1\end{array}\right)$}.
The overall normalization 
should be defined such that for $|x| \gg 1$ they represent D4-branes, 
namely $\ket{\phi} \sim \ket{p=0}$ up to a gauge transformation.
This is a normalization similar to $\phi_a(x)$ in the inverse ADHM
construction. 

Knowing the zeromode expressions, we proceed to get the information on
the surviving D-branes. There are two kinds of D-branes surviving,
corresponding to the fact that we have normalizable and non-normalizable
zeromodes of the tachyon field: the Chan-Paton state $|\psi\rangle$
signals the surviving $k$ D0-branes, and $|\phi\rangle$ shows the
creation of the $N$ D4-branes.\footnote{Note that each non-normalizable
zeromode doesn't correspond to a single D4-brane. In the correspondence
the spinor structure of the SU(2) indices doesn't count.
} 

The location of the surviving $k$ D0-branes is easily found by taking
the expectation value of the original scalar field $\Phi_\mu =
\hat{x}_\mu$, 
\begin{eqnarray}
 X_\mu = \langle \psi |\Phi_\mu |\psi\rangle
= \int \! d^4x 
\langle\psi|x\rangle x_\mu
\langle x | \psi \rangle
\end{eqnarray} 
which is in fact one of the the inverse ADHM formulas, 
(\ref{inverseADHM1}).

Another ADHM data $S$ should be seen from the
D0-D4-string. Remember that the D0-D4-string is encoded in the tachyon
field in the ADHM construction in Sec.~\ref{secidea}.
Since we want the D0-D4-string, what we need is the matrix transition
element of the tachyon between the Chan-Paton states representing the 
D0-branes and the D4-branes. In fact, 
the matrix element of the normalizable and the non-normalizable zero
modes gives 
\begin{eqnarray}
 2\pi uS = i\langle \psi | T^\dagger | \phi\rangle
= u\int \! d^4x 
\langle\psi|x\rangle D
\langle x | \phi \rangle
\label{invs}
\end{eqnarray}
which is nothing but another inverse ADHM formula, (\ref{inverseADHM2}).
Note that for the matrix element of $T$, we have no
normalizable zeromode with positive chirality, and thus the 
expectation value vanishes.
(One might think that
$\langle \phi | T | \psi \rangle=0$ implies 
$\langle \psi | T^\dagger | \phi\rangle =0$, thus contradicts
(\ref{invs}). 
However, strictly speaking, the non-normalizable modes does not 
reside in the Hilbert space. In appendix \ref{appA}, we will justify
(\ref{invs}) by compactifying the $R^4$ to $S^4$ and then 
taking the decompactification limit.)

In this subsection we have derived the inverse ADHM construction from
the D-brane ascent relation in the tachyon condensation.
The important point here is that we have two kinds of D-branes surviving 
the tachyon condensation, and the normalizability of the zeromodes
directly corresponds to the dimensionality of the remaining D-branes.

\subsection{Direct completeness in terms of D-branes}
\label{sec4-2}

The completeness basically means that the ADHM data appearing in
(\ref{diraczero}) is identical with the data obtained by the inverse
ADHM construction (\ref{inverseADHM1}) and (\ref{inverseADHM2}),
once in the inverse ADHM construction one uses the gauge fields 
derived by the ADHM construction. We have already given the 
D-brane realization of these constructions. In this subsection we
further provide a direct way how we can see that those data are
the same, by using again a tachyon condensation. 

The ADHM construction is realized as representing the D0-D4 system by
the D4-anti-D4-branes, while the inverse ADHM construction uses the
D0-anti-D0-branes. Therefore a direct relation between these 
should be seen by representing the D4-anti-D4-branes by infinite number
of the D0-anti-D0-branes. See Fig.~\ref{fig3}.
The way to construct a single D4-brane out of infinite number of
D0-anti-D0-branes is already described in the previous 
subsection, so we just do the same for all the D4-anti-D4-branes. 
Then the resultant D0-anti-D0-brane configuration is as follows.
We have the transverse scalar field $\Phi_\mu^{\rm D0} = 
\Phi_\mu^{\rm antiD0}=\hat{x}_\mu$ as before, as well as 
the tachyon profile 
\begin{eqnarray}
T=\lim_{v\rightarrow\infty}
\bordermatrix{
& \overbrace{\hspace{25mm}}^{2N\infty} 
& \overbrace{\hspace{25mm}}^{4k\infty} 
& \overbrace{\hspace{25mm}}^{4k\infty} 
\cr\nonumber\\[-8mm]
& v \hat{p}_\mu\otimes\unit_{N}\otimes e_\mu^\dagger
& 0 & \unit_\infty\otimes uS \otimes \unit_2 \cr
& 0 & v \hat{p}_\mu\otimes\unit_{2k}\otimes e_\mu^\dagger
& \hspace{4mm} t^\dagger(\hat{x})\hspace{3mm} \otimes \unit_2 \cr
& \unit_\infty\otimes uS^\dagger \otimes \unit_2 
& \hspace{4mm}t(\hat{x}) \hspace{4mm}\otimes \unit_2 
& v \hat{p}_\mu\otimes\unit_{2k}\otimes e_\mu
\cr}
\begin{array}{c}
\bigm\}
{\scriptstyle 2N\infty}
 \\
\bigm\}
{\scriptstyle 4k\infty}
\\
\bigm\}
{\scriptstyle 4k\infty}
\end{array}
\ .
\label{t1}
\end{eqnarray}
The entries including $v$ give rise to the D4-branes and the
anti-D4-branes in the limit $v\to\infty$. 
The upper-left $(2N+4k)\infty \times (2N+4k)\infty$ matrix corresponds
to the $(N+2k)$ D4-branes, while the lower-right part is for the $2k$
anti-D4-branes.  The tachyon configuration (\ref{genericT}) in the
resulting D4-anti-D4-brane appears in the off-diagonal part of the 
total tachyon matrix.\footnote{
Here we identified the off-diagonal elements in (\ref{t1}), 
$t$ and $S^\dagger$, as the tachyon of the D4-anti-D4-branes. 
We easily see that this identification is correct, by
using the Gamma matrix representation when the number of the D4-branes
is the same as that of the anti-D4-branes \cite{Asakawa:2002ui}.
Note that the $T^\dagger$ contains $t,t^\dagger,S$ and $S^\dagger$, so
the chirality operators are different in the D4-brane and the D0-brane
pictures. Actually, an oriented open string connecting a D4-brane and an
anti-D4brane 
is composed of the ones connecting D0-branes and anti-D0-branes
with both orientations.   
}
So if we take $v\to\infty$ limit first, then
we end up with the D4-anti-D4-brane configuration and goes back to 
the starting point of Fig.~\ref{fig1}, {\it i.e.}~the D0-brane point of
view for the ADHM construction.

\FIGURE[t]{
\begin{minipage}{7.7cm}
\begin{center}
\includegraphics[width=7cm]{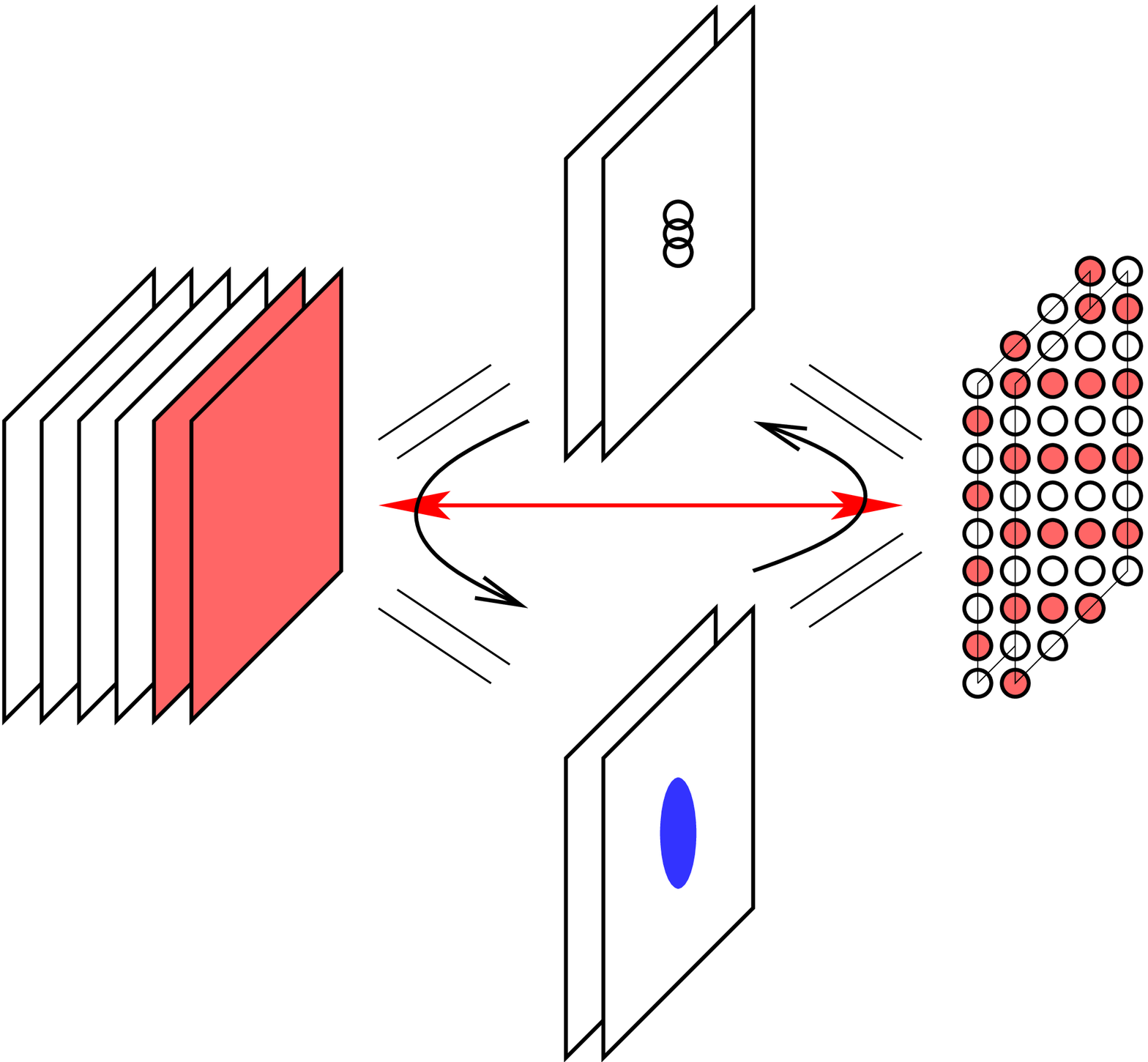}
\caption{The D-brane realization of the completeness. The completeness
and the uniqueness are
equivalent to the fact that this circle with four corners 
is in fact closed. The horizontal arrow is a short-cut, which is
provided in D-brane language and is a proof of the completeness.
}
\label{fig3}
\end{center}
\end{minipage}
}
On the other hand, we can take $u\to\infty$ limit first.
Then, we can use the same gauge transformation $U_1$ and $U_2$
in Sec.~\ref{secder} (but the argument $x$ replaced with $\hat{x}$) 
to diagonalize the $S$ and $t$ part of the matrix (\ref{t1}). With this
gauge transformation, the upper-left $\hat{p}_\mu$ is transformed 
by $U_1$ to $\hat{p}_\mu-i U_1^\dagger \p_\mu U_1$.
Note that, the part of the Chan-Paton indices which represent
the $2k$ pairs of D4-anti-D4-branes 
will drop by the the tachyon condensation in the limit, and therefore
only the upper-left corner survives.
Finally we get the D0-anti-D0 system with
(\ref{tdirac}) where $A_\mu(x)$ is given by the ADHM construction
from the ADHM data $S,X$ appearing in (\ref{t1}).
This means that the D-brane system considered in Sec.~\ref{secder} 
and Sec.~\ref{secderin} are
indeed the same.\footnote{In particular, if we use the completeness
found in \cite{Corrigan:1983sv}, we conclude that the hypermultiplet $S$
appearing in (\ref{genericT}) is identical with $S$ in (\ref{invs}).}

We shall proceed to show the ``direct'' completeness hidden in the big
tachyon matrix (\ref{t1}). 
We consider the same limit as above, $u\to\infty$ first and then
$v\to\infty$, but look at only a part of the tachyon matrix 
(\ref{t1}) ---  the condensation of the lower-right
$4k\infty\times4k\infty$ corner, and diagonalize it first. 
In the limit $u \to\infty$, it is easy to show that
there appear $k$ zeromodes in this part of the tachyon matrix. These
should correspond to the remaining $k$ D0-branes.
When $X_\mu$ in $t$  is simultaneously diagonalizable, 
it is obvious that we get $X_\mu$ as the location of the 
resulting $k$ D0-branes. On the other hand, in the inverse ADHM
construction of Sec.~\ref{secderin}, the data $X_\mu$ is given by the 
location of the $k$ D0-branes, so we are dealing with the same physical
quantity here. 
Therefore we could directly show that $X_\mu$ in the tachyon profile
of (\ref{genericT}) is identical with that of (\ref{locsurd0}). 
This is the completeness.

Let us also derive the completeness for $S$. For simplicity we put
$X_\mu=0$ in the following. After this ``partial'' tachyon
condensation, we may neglect the vanishing pairs of the D0-branes
and anti-D0-branes and deal with only the surviving $k$ D0-branes.  
Using this new and reduced number of basis, one can show that the above
matrix can be represented as\footnote{
Here we demonstrate how we obtained 
the matrix element $vS|x=0\rangle$ in (\ref{resultt1}) briefly. The 
$4k\infty\times 4k\infty$ corner of the matrix (\ref{t1}) simply states
a sequence of the tachyon condensation, $4k\infty$ D0-anti-D0 $\to$ 
$2k$ D4-anti-D4
$\to$ $k$ D0. To avoid the complicated matrix structure of (\ref{t1}),
we consider a simplified sequence $2\infty$ D0-anti-D0 $\to$
D2-anti-D2 $\to$ D0, whose tachyon matrix $\tilde{t}$ and its
normalizable zeromode $\tilde{\psi}(x) = \langle x|\tilde{\psi}\rangle$ 
are given by
\begin{eqnarray}
\tilde{t}= \left(
\begin{array}{cc}
v(\hat{p}_1+i\hat{p}_2) & u(\hat{x}_2-i\hat{x}_1)\\
u(-\hat{x}_2-i\hat{x}_1) & v(\hat{p}_1-i\hat{p}_2)
\end{array}
\right), \quad 
\tilde{\psi}(x) = 
\left(
\begin{array}{c}
1 \\ 1
\end{array}
\right)
\sqrt{\frac{u}{2v}}
\exp\left[-\frac{u}{2\pi v}(x_1^2 + x_2^2)\right] \ .
\nonumber
\end{eqnarray} 
The zeromode wave function becomes $\sqrt{\delta(x)}$ in the
$u\to\infty$ limit. An analog of this wave function in our 
precise 4 dimensional
case for the part of (\ref{t1}) is given by $\langle
x|\tilde{\psi}\rangle = 
\tilde{\psi}_0 (u/2\pi v)\exp[-(u/2v)r^2]$ for $k=1$, where 
$r^2\equiv x_1^2+x_2^2+x_3^2+x_4^2$ and 
$\tilde{\psi}_0\equiv (i,0,0,i,1,0,0,1)^{\rm T}.$
Then, the matrix element which we want to evaluate,
among the $2N\infty\times 8k\infty$ upper-right corner of the
matrix (\ref{t1}), is just 
\begin{eqnarray}
u\left(
\begin{array}{cccccccc}
0&0&0&0&S_1&S_2 & 0 & 0 \\
0&0&0&0&0&0&S_1&S_2 \\
\end{array}
\right)
|\psi\rangle
&&=
\left(
\begin{array}{c}
S_1 \\S_2
\end{array}
\right)
 \int\! d^4x\; |x\rangle u 
\frac{u}{2\pi v}\exp\left[-\frac{u}{2v}r^2\right]
\nonumber \\
&&\hspace{-30mm}= 
\left(
\begin{array}{c}
S_1 \\S_2
\end{array}
\right)
\int\! d^4x\; |x\rangle 2\pi v
\left(\frac{u}{2\pi v}\right)^2
\exp\left[-\frac{u}{2 v}r^2\right]
\stackrel{u\to\infty}{\to}
2\pi vS|x=0\rangle \ .
\nonumber
\end{eqnarray}
Although the procedure of diagonalizing a part of the tachyon matrix
(\ref{t1}) first seems not appropriate, the resultant reduced tachyon
matrix (\ref{resultt1}) has no dependence on $u$ that justifies the
partial diagonalization.
} 
\begin{eqnarray}
T=\lim_{v\rightarrow\infty}
\bordermatrix{
& \overbrace{\hspace{25mm}}^{2N\infty} 
& \overbrace{\hspace{20mm}}^{k} 
\cr\nonumber\\[-8mm]
& v \hat{p}_\mu\otimes\unit_{N}\otimes e_\mu
& 2\pi vS|x=0\rangle 
\cr}
\Bigm\}
{\scriptstyle 2N\infty}
\ .
\label{resultt1}
\end{eqnarray}
We see that ADHM data $S$ is appearing in the tachyon matrix,
in such a way that it is a matrix element of $k$ normalizable modes
and $N$ non-normalizable modes.  
This shows that $S$ is the one given in (\ref{invs}), and the
completeness is proven. In the matrix, $|x=0\rangle$ state is
appearing as a coefficient of $S$, which reflects the fact 
that the normalizable zeromode wave functions are localized at 
the location of the $k$ D0-branes. 
Note that it appears as $2\pi vS$ which
is the correct normalization in view of (\ref{tdirac}) and (\ref{invs})
with replacing $u$ by $v$ there.

\section{Conclusions and Discussions}
\label{sec:5}
\setcounter{footnote}{0}

In this paper, we have derived the ADHM construction of instantons in
string theory. The ADHM procedures appear as a selection process of
remaining D-branes in the tachyon condensation which unifies the
D0-branes and D4-branes. The physical meaning of the ADHM procedures are
found as follows:  
\begin{itemize}
 \item The Dirac operator (\ref{diraczero}) in the ADHM construction is
       the tachyon connecting $N+2k$ D4-branes and $2k$ anti-D4-branes.  
\item The zeromodes of the Dirac operator (\ref{diraczero}) is the
       Chan-Paton wave function of the D4-branes surviving the tachyon 
       condensation. 
\item The ADHM formula (\ref{adhmformula}) is the connection induced by
       the basis change of the Chan-Paton space, looked by the remaining
       D4-branes. It can be viewed also as a Berry's connection on the
       boundary state. 
\end{itemize}
For the inverse ADHM construction, we used the D-brane ascent relation
for the relevant tachyon condensation, and the inverse ADHM formulas
(\ref{inverseADHM1}) (\ref{inverseADHM2}) turned out to be the vacuum
expectation values of the Higgs and the tachyon fields of the system of
infinite number of D0-anti-D0-branes. We have demonstrated that the
completeness can be shown easily in the D-brane setup, and the
derivation of the ADHM construction in noncommutative space was given.

As emphasized in the introduction, the equivalence of the gauge
configurations on the D4-branes and the D0-D4 bound state is quite
nontrivial. Let us say more concretely on this by looking at the 
low energy limits of the two descriptions. For simplicity we consider a
single instanton in SU(2) Yang-Mills theory. The size of the instanton
$\rho$ is proportional to $S$. If we have normalized $S$ such that it
has a scale of length, terms of higher order in $S$ would dominate the
effective action of the D0-branes in the low energy limit 
$\alpha'\to 0$ with $S$ kept finite, and
this means it would not be an appropriate action in the limit. 
Thus naturally $S$ has 
mass dimension one (or at least positive mass dimension), 
and so $\rho=\alpha' S$. The instanton picture is
natural in the $\alpha' \rightarrow 0$ limit with finite instanton size
since there the Yang-Mills action is trustable. On the other hand, the
D0-brane picture, {\it i.e.}~the action using $S$ and $X$, is valid and
natural in the zero slope limit with $S$ kept finite. Therefore these
two pictures reside in different regimes of the validity. 

Nevertheless, we have shown that these two pictures are equivalent
at off-shell. On-shell configurations is given by imposing the
supersymmetry conditions on the boundary state, and in our formalism,
at the opposite end points of the moduli space evidently we can get
the familiar BPS conditions by imposing the supersymmetry conditions:
at $\rho\ll\sqrt{\alpha'}$ we obtain the ADHM equation for $S$ and
$X^\mu$, while at $\rho\gg\sqrt{\alpha'}$ we obtain the instanton
equation. Let us consider the former region where the description by 
the ADHM data is natural. We have shown that the boundary state
at this parameter region is equivalent to the D4-brane boundary state
with the instanton gauge field constructed by the ADHM construction.
But in this latter description the equations of motion for the gauge
field is quite complicated with all order $\alpha'$ corrections. 
This proves that the small instanton configuration of sub-stringy
size is protected against the $\alpha'$ corrections. 

In the middle of the moduli space, since we don't know how the
supersymmetry condition works explicitly, we cannot give any concrete 
result. However, at the both ends of the moduli space, 
$\rho\gg \sqrt{\alpha'}$  and $\rho\ll\sqrt{\alpha'}$, the self-dual  
configuration solves  the BPS equation.
Thus we reach the conjecture stating that {\it self-dual
configurations with arbitrary size of the instanton solve the 
equations of motion of Yang-Mills field corrected in all order in
$\alpha'$in string theory.} In other words, self-dual configurations are 
solutions of non-Abelian Born-Infeld theory with higher derivative
corrections in string theory.

As an affirmative evidence for this conjecture, we note the following
fact. The leading $\alpha'$ correction to the Yang-Mills action on the
D4-branes \cite{Tseytlin:1986ti} is given by the first nontrivial terms
in the expansion of the non-Abelian Born-Infeld action defined with the
symmetrized trace \cite{Tseytlin:1997cs}. In fact, it has been shown
that the self-dual configuration solves the equations of motion of that
theory \cite{Hashimoto:1997px}. So this is the evidence for the
conjecture. At the next order $(\alpha')^3$, it has been known that the
corrections differ from the terms of the action given by the
symmetrized trace \cite{Hashimoto:1997gm}, and their explicit
expression was computed in \cite{Bilal:2001hb,Koerber:2001uu} and has
been successfully tested in \cite{Koerber:2001hk}. Even at this 
$(\alpha')^3$ order
the self-dual configuration solves the equations of motion \cite{KS}. 
(At this order BPS equations in higher dimensions generically have
correction terms, but only in 4 dimensions they vanish.) It
would be interesting to check that the correction terms of the order
$(\alpha')^4$ computed in \cite{Bilal:2001hb,Koerber:2002zb} (and tested
in \cite{Sevrin:2003vs}) may not modify 
the self-dual configuration. 
Note that this is the evidence in the $\alpha'$
expansions of the instantons, and in our paper we give the evidence for
the opposite side of the moduli space of the instanton, that is, small
instantons. These together suggests strongly that the conjecture is
true.\footnote{There is a subtlety concerning
field redefinitions. Since the self-dual equation is not invariant
under some field redefinitions, this conjecture is true only
for some particular definition of fields, although
at the end points of the moduli space the field redefinitions
do not change the BPS conditions because the redefinitions are
$\alpha'$ corrections.
This field redefinition subtlety is related to the choice of 
the regularization of the world sheet theory 
in the BSFT or the boudandary state formalism.} 

In this regard, the results of the present paper ensures that D-brane
techniques for field theory solitons are trustable, in spite of the
difference in the validity of the regions in switching from one
description to the other. The implementation of the solitons by the
tachyon condensation provides a new understanding of the mysterious ADHM
/ Nahm constructions, and opens up new possibilities to view various
other solitons in a different and unified manner. 

Several discussions related to the results of this paper are in order.
\begin{itemize}
 \item Atiyah-Singer index theorem \cite{Atiyah:1968mp}. The index
       theorem is obtained in K-matrix theory \cite{Asakawa:2002ui}. The
       physical equivalence between D$p$-branes and pairs of
       D0-anti-D0-brane can be considered as a generalization of the
       index theorem. (It is related to the topological and analytic
       K-homologies, the KK-theory, the family index theorem and the
       Connes's spectral triple.) There the important ingredient is the
       boundary perturbation, which can be expressed as a quantum
       mechanical partition function. Evaluating it for D0-brane charge,
       namely taking the overlap of the boundary state and the
       Ramond-Ramond state, we
       have the index theorem. The ADHM construction has close relation
       to the index theorem, however, it is not just topological. As we
       have seen in Sec.~\ref{secderin}, the field profile
       (\ref{tdirac}) gives the  ADHM construction in D0-anti-D0-branes
       and the corresponding boundary perturbation is the same one used  
       to show the index theorem.
       Thus we can say that the boundary
       perturbation unifies those. The D4-brane (or a topological
       picture) is obtained in a path-integral representation 
       of the quantum mechanical partition function in 
       the boundary perturbation
       while the
       D0-brane (or an analytic picture) is obtained in an operator
       formalism of it. 
\item Evaluation of the action of the BSFT. In
       this paper we presented the off-shell boundary state, and in
       principle the overlap $\langle 0 | B \rangle$ provides the
       worldsheet partition function which is the boundary superstring
       field theory action. For the tachyon configuration given in this
       paper, it would be possible to compute this action explicitly at
       least as a perturbation in terms of $S$. The resultant equations
       of motion should be consistent with the ADHM equation. This
       explicit check and possible all-order computation in $S$ would
       help the understanding of the BPS nature in the middle of the
       instanton moduli space. 
\item Octonionic instantons \cite{Fubini:1985jm} / higher dimensional
       generalizations 
       of the ADHM construction. A generalized ADHM construction in $4n$
       dimensions has been proposed \cite{Corrigan:1984si}, while it is
       not clear how this can be embedded in string theory. At least for
       $n=2$ ($d=8$), it might be related to D0-D8 bound states
       \cite{Papadopoulos:1997dg} for which one needs D8-anti-D8-branes
       and the appropriate tachyons analogous to our setup. We can
       expect that their noncommutative generalizations 
       \cite{Ohta:2001dh} may follow as in the present paper.
\item We can repeat what we have done in this paper in type I string
       theory instead of type II string theory. Then we will have the
       ADHM construction for the SO or Sp gauge theory. It will be
       interesting to generalize our method to the torus case, namely
       the Nahm transformation. Another interesting extension of the
       D4-D0 bound state is the fuzzy funnel \cite{Constable:2001ag}
       which is an intersecting D1-D5 system (the intersecting D1-D3
       system has been analyzed in our previous paper
       \cite{Hashimoto:2005yy} as the Nahm's construction of 
       monopoles).       
       In this case, there is no supersymmetry, however, our method
       might work because we have not used the supersymmetry
       explicitly.

\end{itemize}

\acknowledgments 

K.~H.~appreciated insightful conversations with
K.~Hori, P.~Koerber, P.~Mukhopadhyay, 
N.~Nekrasov, M.~Nitta, A.~Sevrin, Y. Tachikawa, D.~Tong and 
A.~Tseytlin.
S.~T.~would like to thank M.~Douglas, K.~Furuuchi, R.~Karp 
and T.~Takayanagi 
for valuable discussions and comments. 
S.~T.~is grateful also to the organizers of 
the Summer Institute String Theory 2005 at Sapporo for 
a stimulating environment.
K.~H.~was supported partly by the National Science Foundation under
Grant No.~PHY99-07949, by the Japan Society for Promotion of Science,
by the Royal Society International Grants, and by the Japan Ministry
of Education, Culture, Sports, Science and Technology.
The work of S.~T.~was supported in part by DOE grant
DE-FG02-96ER40949.


\appendix

\section{Note on Non-Normalizable Zeromodes}
\label{appA}

In this appendix we consider a consistent definition of the Dirac
operator to resolve the problem mentioned below the equation
(\ref{invs}). The meaning of the ADHM formula (\ref{inverseADHM2}) will
become clearer. 

We define the Dirac operator $\cal D \equiv \gamma_\mu D_\mu$ which is
now the usual definition in the $4\times 4$ matrix form, 
and the chirality operator $\Gamma^5$. Let us consider the
non-normalizable zeromode $\phi^+$ which appeared in
Sec.~\ref{secderin}. 
The superscript indicates the chirality of the spinor.
Then, because of the equation $(D_\mu D_\mu \otimes \unit_2) \phi^+=0$, 
$v^-\equiv D \phi^+$ 
is a zeromode of $D^\dagger$, {\it i.e.}~$D^\dagger v^-=0$.
Due to the asymptotic behavior of $\phi^+$,
we can see that $v^-$ is normalizable and then can be written as 
a linear combination of $\psi^-$.
Therefore, we obtain a simple relation $v^-=S \psi^-$
where $S$ is the $N \times 2k$ matrix which 
acts on the spinor index and 
also on the index labelling the $k$ normalizable zeromodes. 

Note that in $R^4$, the zeromode of ${\cal D}^2$
is not necessarily a zeromode of ${\cal D}$. 
Thus if we include the non-normalizable modes,
${\cal D}$ is not Hermitian and
it is unclear how to define ${\cal D}$. This is the problem which we
mentioned below the equation (\ref{invs}).
To overcome this problem,
we consider an $S^4$ and taking a limit of large radius to $R^4$. In
this way,  
we can justify the use of the non-normalizable modes, 
and ${\cal D}$ is now defined as a Hermitian operator.
In a compact space, an eigen state of ${\cal D}^2$, 
${\cal D}^2 \ket{m^2}=m^2 \ket{m^2}$, 
can be decomposed as $\ket{m^2}=\ket{m} +\ket{-m}$ for $m \neq 0$
where ${\cal D}\ket{\pm m}=\pm m \ket{\pm m}$ and
$\Gamma^5 \ket{m}=\ket{-m}$.
Then any zeromode of ${\cal D}^2$ is a zeromode of ${\cal D}$.
This seems to be contradicting the $R^4$ case.
However, we will see it is not, by looking at the large radius limit
carefully below. 

Consider an $S^4$ with a very large radius.
Then there are eigenmodes whose eigenvalues are zero or very close to
zero which correspond to 
$\psi^-$ and $\phi^+$ in the large radius limit.
Suppose that $\eta$ which is not a zeromode corresponds to the $\phi^+$.
Then we have
\begin{eqnarray}
({\cal D} +m)\eta=0 \ .
\end{eqnarray}
And this is written in the chiral decomposed form as 
\begin{eqnarray}
 \left(
\begin{array}{cc}
m\unit_2 &D \\
D^\dagger & m\unit_2
\end{array}
\right)
\left(
\begin{array}{c}
\eta^- \\ \eta^+
\end{array}
\right)
=0 \ ,
\end{eqnarray}
where in the limit to $R^4$ the mass parameter $m$ is taken to $0$.
This equation means 
\begin{eqnarray}
 (D^\dagger D - m^2) \eta^+=0 \ , \quad 
\eta^- = -\frac{1}{m} D \eta^+ \ .
\end{eqnarray}
Since $\phi^+$ is non-normalizable and 
we supposed $\eta^+ =C \phi^+$ in the limit,
the normalization constant $C$ appearing here should go to zero in the
limit. (Otherwise the above equation wouldn't make sense.)
Thus, in the large radius limit $m\to 0$, $\frac{C}{m}$ will be kept
finite  
and $\eta^-$ should be a linear combination of $\psi^-$
since it is normalizable.
In this way we can have a consistent result,
\begin{eqnarray}
D^\dagger D \phi^+=0 \ , \quad 
S \psi^- = D \phi^+ \ .
\end{eqnarray}
Here the important point is that
the usual normalization of the state in $S^4$ is different from 
the plane wave normalization of the state in $R^4$.

\newcommand{\J}[4]{{\it #1} {\bf #2} (#3) #4}
\newcommand{\andJ}[3]{{\bf #1} (#2) #3}
\newcommand{\AP}{Ann.\ Phys.\ (N.Y.)}
\newcommand{\MPL}{Mod.\ Phys.\ Lett.}
\newcommand{\NP}{Nucl.\ Phys.}
\newcommand{\PL}{Phys.\ Lett.}
\newcommand{\PR}{ Phys.\ Rev.}
\newcommand{\PRL}{Phys.\ Rev.\ Lett.}
\newcommand{\PTP}{Prog.\ Theor.\ Phys.}
\newcommand{\hep}[1]{{\tt hep-th/{#1}}}

\end{document}